\documentstyle[12pt]{article}

\def\hybrid{\topmargin -20pt   \oddsidemargin 0pt
      \headheight 0pt   \headsep 0pt
      \textwidth 6.25in % A4 paper
      \textheight 9.5in % A4 paper
      \marginparwidth .875in
      \parskip 5pt plus 1pt   \jot = 1.5ex}

\hybrid

\def\x{\times}

\def\o+{\oplus}
\def\ra{\rightarrow}
\def\lra{\longrightarrow}

\def\leria{\leftrightarrow}
\def\llra{\longleftrightarrow}

\def\beqa{\begin{eqnarray}}
\def\eeqa{\end{eqnarray}}
\def\exp{\mbox{exp}}

\def\back{\backslash}
\def\om{\omega}
\def\th{\theta}
\def\al{\alpha}
\def\be{\beta}
\def\ga{\gamma}
\def\de{\delta}

\def\ka{\kappa}
\def\la{\lambda}
\def\si{\sigma}
\def\Y{\Upsilon}
\def\pa{\partial}
\def\sign{\mbox{sign}}
\def\det{\mbox{det}}

\def\vol{\mbox{vol}}

\def\Im{\mbox{Im}}
\def\Re{\mbox{Re}}

\def\we{\wedge}

\def\u{\tilde{u}}
\def\A{{\cal A}}

\def\C{{\cal C}}
\def\D{{\cal D}}

\def\F{{\cal F}}

\def\L{{\cal L}}
\def\M{{\cal M}}

\def\O{{\cal O}}

\def\U{{\cal U}}

\def\De{\Delta}
\def\Ga{\Gamma}
\def\La{\Lambda}
\def\Om{\Omega}

\newcommand{\resetcounter}{\setcounter{equation}{0}}

\begin{document}

\begin{titlepage}

\begin{flushright}
\hfill{HU-EP-02/59} \\
\hfill{hep-th/0212233}
\end{flushright}

\vspace{15pt}

\vspace{1cm}
\begin{center}

{\Large Superpotential of the $M$-theory conifold}\\
\vspace{.2cm}
{\Large  and type IIA string theory}

\vspace{30pt}

{Gottfried Curio}

%\vspace{10pt}

{\it  Humboldt Universit\"at zu Berlin,
Institut f\"ur Physik,\\
Invalidenstrasse 110, 10115 Berlin,
Germany}
\vspace{20pt}

\end{center}

The membrane instanton superpotential for
$M$-theory on the
$G_2$ holonomy manifold given by
the cone on ${\bf S^3}\x {\bf S^3}$
is given by the dilogarithm and has
Heisenberg monodromy group
in the quantum moduli space.
We compare this to a Heisenberg group action
on the type IIA hypermultiplet moduli space
for the universal hypermultiplet,
to metric corrections from membrane instantons
related to a twisted dilogarithm for the deformed conifold
and to a flat bundle related to a conifold period,
the Heisenberg group and the dilogarithm
appearing in five-dimensional Seiberg/Witten theory.

\vspace{1cm}

\end{titlepage}

\section{\label{introduction}\large {\bf Introduction}}

The goal of this paper,
which may be read as a companion paper to [\ref{C}],
is to put into perspective the occurrence of
the dilogarithm and the Heisenberg group action
for the $M$-theory conifold
via comparison to analogous phenomena
in type IIA string theory
(which are not just treating the identical problem
of describing the superpotential [\ref{AV}]).
%Let us recall the $M$-theory case first.

In the conifold transition between Calabi-Yau manifolds in type II
string theory an ${\bf S^3}$ is exchanged with an ${\bf S^2}$.
When lifted to $M$-theory [\ref{AW}], [\ref{AMV}]
the resulting geometries become symmetrical:
the two small resolutions on the one side of the transition
become ${\bf S^3}$'s as well,
Hopf fibred by the $M$-theory circle ${\bf S^1_{11}}$.
This corresponds in type IIA to having
one unit of RR flux on the ${\bf S^2}$;
on the deformed conifold side
one has a $D6$ brane wrapped on the ${\bf S^3}$.
$M$-theory of the corresponding non-compact
$G_2$ holonomy manifold (with covariant constant three-form $\Y$)
is a deformation $X_7$ of the cone over
${\bf S^3}\x{\bf S^3}=SU(2)^3/SU(2)_{diag}$.
It is topologically ${\bf R^4}\x {\bf S^3_Q}$
(where ${\bf R^4}$ comes from the 'filled in' ${\bf S^3_D}$)
and comes in three triality equivalent versions $X_1, X_2, X_3$
[\ref{AW}], [\ref{AV}], [\ref{C}].

The superpotential given by the multi-cover membrane instantons
is given by the dilogarithm [\ref{AV}], [\ref{C}]
of the modulus $u=e^{i\Phi}$ where $\Phi=\int_Q C+i\Y$
\beqa
\label{Li}
Li(u)=\sum_{n=1}\frac{u^n}{n^2}=-\int_0^u \log(1-u) d\log u
\eeqa
The multi-valuedness is described by the
monodromy representation of the fundamental group
$\pi_1 ({\bf P^1} \backslash  \{ 0,1,\infty \})$.
The relevant local system is given by a bundle,
flat with respect to a suitable connection.
In the case of the logarithm the
monodromy of $\log z$ around $z=0$
is described by adding integers (times $2\pi i$)
what is captured by the monodromy matrix
$\left(\tiny{\begin{array}{cc}1&2\pi i\\0&1\end{array}}\right)$
acting on the two-vector $(\log z , 1)^t$;
%the monodromy group is given by ${\cal U}_{\bf Z}\hra {\cal U}_{\bf C}$
%where ${\cal U}$ denotes the upper triangular group
%$\scriptsize{\left(\begin{array}{cc}1&*\\0&1\end{array}\right)}\subset
%Sl(2)$ (the embedding of ${\cal U}_{\bf Z}$ in ${\cal U}_{\bf C}$ may
%include the factor of $2\pi i$).
the generalisation in the case of the dilogarithm
involves the Heisenberg group of upper triangular $3\x 3$ matrices
acting on the three-vector $c_3$.
The Heisenberg bundle gives for the monodromy at infinity [\ref{C}]
\beqa
c_3=\small{\left( \begin{array}{c}Li(y)\\ \log y \\ 1 \end{array} \right)}
\;\;\;\; , \;\;\;\;
M(l_{\infty})=  \small{\left( \begin{array}{ccc}
                  1 &     2\pi i  &   0        \\
                  0 &      1      &  -2\pi i   \\
                  0 &      0      &   1
                  \end{array} \right)  }
\eeqa

Now let us compare to the type IIA results.
In the case of the (deformed) conifold one has
a relation between the quaternionic type IIA modulus
and the $M$-theory modulus.
The metric on the hypermultiplet moduli space
is corrected by membrane instantons [\ref{OV}]
wrapping the ${\bf S^3}$
and, taking into account normalizations [\ref{GG}],
(a twisted version of)
$Li$ of the corresponding ${\bf S^3}$ modulus
occurs in a suitable hyperk\"ahler limit of the quaternionic moduli
space geometry (where the details of the embedding of the local
conifold $T^* {\bf S^3}$ in a specific global Calabi-Yau become
irrelevant).
Here the  $N=1$ superpotential $W=Li(u)$ corresponds to the
$N=2$ metric corrections
just as one and the same Schwinger computation has different
physical interpretations in $N=0 \, , 1 \, , 2$ (cf. below).

Going beyond that limit but keeping the aimed at universality
we further compare with
the moduli space of the universal hypermultiplet
(cf. [\ref{BB}] - [\ref{Ketov}])
and a Heisenberg action on it. In the $M$-theory
conifold the quantities $\int_{{\bf S^3_{Q/D}}} *_7 G$
(analogous to certain type IIA membrane charges, cf. sect. \ref{varia})
accompany
in the scalar potential term
$\int_D C \cdot \int_Q *_7 G$
(the imaginary parts of)
$\log \eta = \log \be u$ and $\log u$
which occupy
%the $L_{12}, L_{23}$
places in the Heisenberg geometry
(\ref{flat section}) corresponding to
the positions in the Heisenberg geometry
(\ref{IIA Heisenberg}) of the type IIA
membrane charge parameters $\be, \ga$.

Our second theme is to compare to
five-dimensional Seiberg/Witten theory
on ${\bf R^4}\x {\bf S^1_R}$
([\ref{Nekrasov}] - [\ref{Eguchi Kanno}]).
There the dilogarithm $Li$
occurs when describing the dual period $A_D$
in the flat bundle of 5D $N=2$ $SU(2)$ field theory.
Note that the 4D period $a_D$ is related to the
conifold\footnote{at the monopole point $\tilde{u}=1$, i.e.
$\Xi_6\sim a_D \sim x_+ \sim \tilde{u}-1$,
by (\ref{locking}), (\ref{periods}), where
$\Xi_6$ is the period related to the $6$-cycle in type IIA
respectively the vanishing ${\bf S^3}$ of the conifold in type IIB}.
In the weak coupling limit
one has for the monodromy at infinity
(in suitably normalized periods
and measuring the quantum scale $\La$ by the
compactification scale $m_R=1/R$ in terms of
the dimensionless quantity $\zeta=\frac{\La}{m_R}$)
\beqa
\small{\left( \begin{array}{c}\tilde{\A_D}\\ \tilde{\A} \\ 1
\end{array} \right)}
=\small{\left( \begin{array}{ccc}
                  1 &  \log \zeta &   0              \\
                  0 &      1      &   0              \\
                  0 &      0      &   1
                  \end{array} \right)   }
\small{\left( \begin{array}{c}Li(x)\\ \log x \\1\end{array}\right)}
\;\;\; , \;\;\;
\tilde{M}_{\infty}=  \small{\left( \begin{array}{ccc}
                  1 &   -2\cdot 2\pi i    & 2\pi i\log \zeta    \\
                  0 &      1      &    2\cdot 2\pi i            \\
                  0 &      0      &   1
                  \end{array} \right)}
\eeqa
Therefore one has that the 5D analogue $\tilde{\A_D}$
of the conifold period $a_D$
is given by the dilogarithm of (the exp of) $\tilde{\A}$
up to a linear modification (cf. [\ref{C}], sect. 4.4)
\beqa
\tilde{\A}_D=Li(e^{\tilde{\A}})+(\log \zeta) \tilde{\A}
\eeqa

We will also remark on relations between
the monodromy at the conifold point
(in a type IIB vector multiplet moduli space language)
and $M$-theory results, and on
relations between the way of derivation
of the membrane instanton superpotential
(\ref{Li}) and an instanton reinterpretation [\ref{KimPage}]
of the occurrence of $Li$
as an imaginary energy density $Li(e^{2\pi i\tau })$
(with $\tau = i\frac{m^2}{2eE}$,
giving the probability density
for the pair creation in a constant electric background $E$)
in the classical non-supersymmetric Schwinger computation.

Note the different meanings of $Li$:
the pair production rate $w$ (related to the vacuum
transition amplitude $P_0$) for $N=0$ (Schwinger),
the disc instanton superpotential $W=F_{0,1}$
(an open topological string amplitude)
respectively the membrane instanton superpotential for $N=1$,
the $R^2$-correction  $F_1$ (a closed topological string amplitude)
for $N=2$ (cf. app. \ref{Schwinger} and footn. \ref{Schwinger footn}),
and the metric correction (\ref{twisted Li}).

This paper is organized as follows.
In {\em Section 2} we compare the $M$-theory situation with
the type IIA hypermultiplet moduli space:
we consider a Heisenberg action on it
for the universal hypermultiplet
and point to a
relation of a twisted dilogarithm
with metric corrections for the deformed conifold,
point to $M$-theory relations reflecting the conifold
monodromy and parallelisms of the $M$-theory treatment
of the membrane instanton superpotential
with an instanton description of
the classical ($N=0$) Schwinger computation.
In {\em Section 3} we describe how $Li$ occurs,
related to the 5D analogue of the conifold period $a_D$,
in the three-dimensional flat bundle of
$N=2$ pure $SU(2)$ gauge theory in 5D.
In {\em Section 4} we collect the observations in an
interpretational discussion.
In the {\em appendix} we recall background for
the 5D Seiberg/Witten theory,
as well as on stringy resp. $M$-theoretic embeddings,
and on the Schwinger computation.

\section{\large\label{varia}{\bf Structures in
the type IIA hypermultiplet moduli space}}
\resetcounter

A hypermultiplet of the effective four-dimensional theory
of the type IIA string on a Calabi-Yau has a quaternionic modulus
consisting of two complex parts. One is given
by the integral of the
holomorphic three-form $\Om$ over a three-cycle $A={\bf S^3}$, say,
the other by the integral of the Ramond three-form $C$ over
$A$ and a dual cycle $B$.
In aiming at a certain universality
different approaches are possible. Either one can localise
and consider the deformed conifold $T^*{\bf S^3}$;
this concerns a hypermultiplet related to a
{\em specific} three-cycle in the Calabi-Yau manifold $X$,
but zooming to the local limit
(hyperk\"ahler reduction of the quaternionic geometry) gives those
results a certain universality (cf. remark 1 below).
This makes them amenable to a comparison
with the other possible approach which treats
the universal hypermultiplet.

\noindent
{\em The conifold transition in type IIA and $M$-theory}

In a type IIA reinterpretation (cf. [\ref{AW}])
one divides by the circle ${\bf S^1_{11}}=U(1)\subset SU(2)_1$
giving for $X_1={\bf R^4}\x {\bf S^3}
=(SU(2)_1 \x {\bf R}^{\geq 0})\x {\bf S^3}$ the type IIA manifold
$({\bf S^2} \x {\bf R}^{\geq 0})\x {\bf S^3}={\bf R^3}\x {\bf S^3}$
with fixed point at the origin, i.e. the deformed conifold
$T^* {\bf S^3}$ with a D6-brane wrapping the zero-section
(the cycle ${\bf S^3_A}$ below). For $X_2$ or $X_3$ one gets
${\bf R^4}\x {\bf S^3}/U(1)={\bf R^4}\x {\bf S^2}$, the two
small resolutions of the conifold together with a unit of RR
two-form flux on ${\bf S^2}$ (as ${\bf S^3}$ is Hopf fibered
by ${\bf S^1_{11}}$ over ${\bf S^2}$).

For the deformed conifold $T^*{\bf S^3}$
one has as relation between the type IIA (quaternionic) hypermultiplet modulus
$Z$ (with complex part $z=\int_Q \Om$, here $Q=A$ is the ${\bf S^3}$)
and the $M$-theory variable $\Phi=\int_Q C+i\Y$.
Let $A$ and $B$ the compact and dual non-compact
three-cycle, respectively: one has $A\simeq Q_1$ but
$B={\bf R^3}={\bf S^3_1}/U(1) \x {\bf R^{\geq 0}}\not\simeq D_1={\bf S^3_1}$,
$({\bf R^4}, D)$ corresponds after the circle-reduction to $(B, {\bf S^2_B})$.
$A=Q$ is a supersymmetric cycle
and $\Re \, \Omega|_A=vol|_{{\bf S^3}}=\Upsilon|_Q$ and
$\int_A\Im \; \Omega =0$. Further one has classically in $M$-theory that
$\int_D C=0$, but $B\not \simeq D$ and
rather $\int_D C = \int_{{\bf S^2}}B^{(2)}
=\int_{{\bf R^3}}H^{(3)}\neq \int_{{\bf R^3}} C$
(the first equality by circle reduction of field and cycle).
One may consider further the fate of the
remaining $N=2$ parameter $\int_B C$ from the perspective of
the additional $D6$ brane on ${\bf S^3_Q}$
which effects the reduction to $N=1$.

A local\footnote{For comments on the somewhat
non-standard technical sense of the
notions 'modulus' and 'superpotential' caused by the non-compactness
of $X_7$, and on the question of having a global coordinate on the
quantum moduli space versus having a first order parameter,
cf. [\ref{C}].}
parameter on the corresponding (quantum) moduli space
in $M$-theory is
given by the membrane instanton amplitude $u=e^{ i \Phi }$
(cf. app. \ref{cone app}).
The membrane instanton superpotential $W(u)$
is given [\ref{C}] by the dilogarithm
(cf. app. \ref{Special functions})
\beqa
\label{Li repeat}
Li(u)=\sum_{n\geq 1}\frac{u^n}{n^2}=-\int_0^u \log(1-u) d\log u
\eeqa
There are two equivalent ways to express the monodromy.
In a {\em vector picture} one assembles $Li$, the ordinary logarithm
and the constants to a three-vector and analytic continuation
along a loop $l_i$ in ${\bf P^1}\backslash \{ 0,1,\infty \}$
(encircling $z=i$) leads to
the monodromy representation
$M: \pi_1 ({\bf P^1} \backslash  \{ 0,1,\infty \}) \ra Gl(3,{\bf C})$
\beqa
\label{monodromy matrices}
c_3=\left( \begin{array}{c}Li(u)\\ \log u \\ 1 \end{array} \right)
\; : \;\;\;\;
M(l_0)=  \left( \begin{array}{ccc}
                  1 &      0      &   0  \\
                  0 &      1      &   2\pi i   \\
                  0 &      0      &   1
                  \end{array} \right)
\;\; , \;\;
M(l_1)=  \left( \begin{array}{ccc}
                  1 &     -2\pi i       &   0  \\
                  0 &      1      &   0  \\
                  0 &      0      &   1
                  \end{array} \right)
\eeqa
Alternatively, in a {\em Heisenberg picture}, consider the complexified
Heisenberg group ${\bf {\cal H}_C}$
of upper triangular complex $3\x 3$ matrices with $1$'s on the diagonal.
%\beqa
%\left( \begin{array}{ccc}
%                  1 &      a      &   c   \\
%                  0 &      1      &   b   \\
%                  0 &      0      &   1
%                  \end{array} \right)
%\buildrel \cong \over \lra
%(a,b \, | \, c)\in {\bf {\cal H}_C}
%\eeqa
Instead of $c_3$ one considers here the expression
(a flat section of a suitable connection [\ref{C}])
(cf. (\ref{Sl2 realisation}))
\beqa
\label{flat section}
\Lambda(u)=\left( \begin{array}{ccc}
                  1 &- \log \be u &   - Li(u)  \\
                  0 &      1      &    \log u  \\
                  0 &      0      &   1
                  \end{array} \right)
\eeqa
and left operation with ${\bf {\cal H}_Z}$ expresses
the multi-valuedness (\ref{multi valued}),
cf. footn. \ref{flat sect footn}.
More precisely one has for the
monodromy along the loops $l_i$ representing left multipliers $h_i$
\beqa
\label{heis mono}
h_0=\left( \begin{array}{ccc}
                  1 &      0      &   0        \\
                  0 &      1      &   2\pi i   \\
                  0 &      0      &   1
                  \end{array} \right)
\;\;\; , \;\;\;
h_1=\left( \begin{array}{ccc}
                  1 &  2\pi i     &   0        \\
                  0 &      1      &   0        \\
                  0 &      0      &   1
                  \end{array} \right)
\eeqa
One has a real-valued non-holomorphic
monodromy-invariant (i.e. single-valued) function
\beqa
\label{inv fct}
\L (u) = \Im \, Li (u) - \Im \, \log \be u \, \Re \, \log u
\eeqa

These results arise as follows.
One has from $(a,b \, | \, c)\ra (x,y)=(e^a, e^b)$ a bundle
over ${\bf C^*_x}\x {\bf C^*_y}$
with fibre\footnote{the entries of ${\bf {\cal H}_Z}$ are actually
from $((2\pi i) {\bf Z}, (2\pi i) {\bf Z} \, | \, (2\pi i)^2 {\bf Z}$);
here $(a,b|c)$ gives the $g_{12}, g_{23}, g_{13}$ entry}
$(2\pi i)^2 {\bf Z}\backslash {\bf C_c}$
(isomorphic to ${\bf C^*}$ via $c \ra S:=e^{c/2\pi i}$)
\beqa
\label{heisenberg bundle}
&{\bf {\cal H}_Z}\backslash {\bf {\cal H}_C}& \nonumber\\
&\downarrow&\nonumber\\
&{\bf C^*_x}\x {\bf C^*_y}
=(2\pi i {\bf Z})^2 \backslash {\bf C^2_{a,b}}&
\eeqa
A pullback gives [\ref{C}] the {\it Heisenberg bundle}
$\underline{{\cal H}}$ over ${\bf P^1}\backslash \{ 0,1,\infty \}$
\beqa
\label{proper heisenberg bundle}
\begin{array}{ccc}
\underline{{\cal H}} & \lra & {\bf {\cal H}_Z}\backslash {\bf {\cal H}_C}  \\
(2\pi i)^2 {\bf Z}\backslash {\bf C} \downarrow
\;\;\;\;\;\;\;\;\;\;\;\;\;\;\;\;\;
& & \downarrow \\
{\bf P^1}\backslash \{ 0,1,\infty \} & \buildrel (1-z,z) \over \lra
& {\bf C^*} \x {\bf C^*}
\end{array}
\eeqa
A section $s$ of $\underline{{\cal H}}$ has the form
$s(z)={\bf {\cal H}_Z}(- \log \be z , \log z \, | \, c)$ and
(\ref{flat section}) is a {\it flat}
section\footnote{\label{flat sect footn}One finds [\ref{C}]
(undoing the fibre identification $c \ra e^{c/2\pi i}=S$) that
the flatness condition $dc=u\, dv$ just expresses the $Li$ integral, and that
the coset takes into account
the multi-valuedness (\ref{multi valued}).}.

\noindent
{\em Heisenberg action for the universal hypermultiplet}\\
\noindent
For the universal hypermultiplet
(cf. [\ref{BB}] - [\ref{Ketov}])
the quaternionic modulus consists
of a complex part given by the four-dimensional
dilaton-axion combination $e^{-2\phi}+2iD$ where the pseudoscalar $D$
arises by dualizing the external (Minkowski) component of $H_3=dB_2$
\beqa
\label{H ext}
H_3&=&e^{4\phi}*_4\Bigl( 2dD +
i (\bar{c} d  c - c d \bar{c}) \Bigr)
%\\
%\label{J al}
%J_{\al}&=&e^{2K}\Bigl(-i d(S - \bar{S})
%+ 2 i( \bar{c}dc-c  d \bar{c}) \Bigr)
\eeqa
together with a complex
$C$-field period $c$
\beqa
\label{C field component}
C_{ijk}=\sqrt{2}\,c\, \Om_{ijk}
\eeqa
where we used the complex field
\beqa
S=e^{-2\phi}+2iD + c\, \bar{c}
\eeqa
and the K\"ahler potential
\beqa
K=-\log (S+\bar{S}-2c\, \bar{c})= 2\phi \;  \; -\log 2
\eeqa
The Lagrange density for the scalar fields takes the form
\beqa
\L &\sim &(d\, \phi)^2 +e^{2\phi}|d\, c|^2
+ \frac{1}{2}e^{4\phi}
\Bigl(2 d\, D + i(\bar {c} d  c - c d \bar{c})\Bigr)^2\nonumber\\
&=&K_{S\bar{S}}\, dSd\bar{S} + K_{S\bar{c}}\, dSd\bar{c}
+ K_{c\bar{S}}\, dcd\bar{S} + K_{c\bar{c}}\, dcd\bar{c}\nonumber\\
=\; ds^2&=&e^{2K}\Bigl(dSd\bar{S} -2c \, dSd\bar{c}
-2\bar{c} \, dcd\bar{S} + 2(S+\bar{S})dcd\bar{c}\Bigr)
\eeqa
The mentioned line element for the space spanned by the parameters
$S$ and $c$ gives the quaternionic space $SU(2,1)/U(2)$. The classical
Lagrangian has eight symmetries (the left $SU(2,1)$ action), four of
which (related to scale transformations and further isometries)
are not preserved in the quantum theory. By contrast from the
remaining four symmetries one gets symmetries even present when
the membrane and fivebrane instantons are taken into account,
broken to their discrete parts [\ref{BB}]. This leads first to
the ${\bf Z_2}$ generated by the involution
\beqa
\label{involution}
\iota: \Re \, c \longleftrightarrow \Im \, c
\eeqa
Then there is a group of isometries associated with the following
shifts ($\al , \be , \ga \in {\bf R}, \; \de = \ga + i \be$; note that,
like the action of $\iota$, this leaves $K$ invariant)
\beqa
\label{IIA Heisenberg}
\left( \begin{array}{c} S \\ c \\ 1  \end{array} \right)
\lra
\left( \begin{array}{ccc}
                  1 &      2\de   &  \de \bar{\de}+i\al   \\
                  0 &      1      &   \bar{\de}           \\
                  0 &      0      &   1
                  \end{array} \right)
\left( \begin{array}{c} S \\ c \\ 1  \end{array} \right)
\eeqa
Note that when just turning on an individual parameter $\al , \be ,
\ga$, leading to corresponding transformations $T_{\al}, T_{\be}, T_{\ga}$,
one has the characteristic relation
$T_{\be}T_{\ga}=T_{\al=4\be\ga}T_{\ga}T_{\be}$.
If $Z$ denotes the corresponding group with {\em integral} entries
(enhanced with $\iota$)
then the moduli space is actually $\M=Z\back SU(2,1)/U(2)$.
Note the invariant expression (cf. (\ref{inv fct}))
\beqa
\label{invariant}
e^{-2\phi}=\frac{1}{2}e^{-K}= \Re \, S - c \, \bar{c}
\eeqa

As in
(\ref{flat section}) one may consider the action not on
the vector $(S,c,1)^t$ but on a Heisenberg group element;
for this consider the rescaled coordinate $e:=\sqrt{2}c$,
$\eta:=\sqrt{2}\de$
\beqa
\label{IIA Heisenberg in e}
\left( \begin{array}{ccc}
                  1 &   \bar{e}   &   S  \\
                  0 &      1      &   e  \\
                  0 &      0      &   1
                  \end{array} \right)
\lra
\left( \begin{array}{ccc}
                  1 &      \eta   &  \frac{\eta \bar{\eta}}{2}+i\al   \\
                  0 &      1      &   \bar{\eta}           \\
                  0 &      0      &   1
                  \end{array} \right)
\left( \begin{array}{ccc}
                  1 &   \bar{e}   &   S  \\
                  0 &      1      &   e  \\
                  0 &      0      &   1
                  \end{array} \right)
\eeqa
with the invariant expression
%\beqa
$2e^{-2\phi}=e^{-K}=  (S+\bar{S}) - e \, \bar{e}$.
%\eeqa

Note further that one has a contribution
to the $dc d\bar{c}$ component of the metric
as given by the following term in the action
(up to the $e^{4\phi}$ normalization factor)
\beqa
\label{metric component}
2\, \Re \, S \, dc \, d\bar{c}
\eeqa

Associated with the three isometries
are classically conserved Noether
currents, first $J_{\al}=2*H_3$
with the corresponding conserved fivebrane charge
$Q_{\al}=\int_{\C_3}* J_{\al}=2 \int_{{\bf S^3}}H_3$
for an euclidean fivebrane wrapping the whole $X$,
being surrounded by an ${\bf S^3}$ in the transversal
non-compact ${\bf R^4}$.
More important for us are the other two currents
\beqa
J_{\be}&=&2ie^Kd(c-\bar{c})+2(c+\bar{c})J_{\al}\nonumber\\
J_{\ga}&=&2e^Kd(c+\bar{c})-2i(c-\bar{c})J_{\al}
\eeqa
where the corresponding conserved charges
\beqa
Q_{\be , \ga}=\int_{\C_3}* J_{\be , \ga}
\eeqa
correspond to membrane charges\footnote{Note that,
very schematically, $J_{\be / \ga}\sim dc+c\cdot
*_4H_3$ and so $*_4J_{\be / \ga}\sim *_4 dc + c \cdot H_3$ and
$(dC)_{\mu ijk}\sim \pa_{\mu}c \, \epsilon_{ijk}$
by (\ref{C field component}), so for this component
$*_{10}(dC)\sim *_4\pa_{\mu}c *_6\epsilon_{ijk}
=*_4\pa_{\mu}c \;\; \epsilon_{\bar{i}\bar{j}\bar{k}}$;
the $c \cdot H_3$ matches as well.}
\beqa
\label{membrane charges}
\int_{\C_6}e^{\phi/2}*_{10}dC + C \we H_3
\eeqa

Compare in the $M$-theory conifold case the quantities
$\int_{{\bf S^3_{Q/D}}}*_7 G$. In the
scalar potential term\footnote{either for a flux or in
a two-point function with an auxiliary $G$ to evaluate $W_{mem}$ [\ref{C}]}
$\int_X |G|^2=\int_B G \cdot \int_Q *_7 G \leria \int_D C \cdot \int_Q *_7 G$
they accompany (the imaginary parts of)
$\log \eta = \log \be u$ and $\log u$
which occupy the $\La_{12}, \La_{23}$ places in (\ref{flat section})
just as do the membrane parameters $\be, \ga$
(which go with $Q_{\be \ga}$)
in (\ref{IIA Heisenberg})
(a ${\bf Z_2}$ exchanges $Q$ and $D$; cf. $\iota$ above).

One may parametrize the coset space $\M_{UH}=SU(2,1)/U(1)$ also
by two complex scalars $z_1$ and $z_2$ in the open
four-ball $|z_1|^2+|z_2|^2<1$ (cf. app. \ref{UH mod sp})
which relate to $S$ and $c$ by
$S = \frac{1-z_1}{1+z_1}\; , \; c = \frac{z_2}{1+z_1}$;
introduction of polar coordinates (\ref{polar coord}) gives then
a $T^2=\La \backslash {\bf C}$
fibration\footnote{the lattice $\La$ occurring here is not to be
confused with the flat section (\ref{flat section})}
over the open quarter-disc $\D$ parametrised by
$z:= |z_1|+i |z_2|$
\beqa
\label{bundle comparison}
\begin{array}{ccc}
\M_{UH} & \;\;\;\;\;\;\;\;\;\;\;\;\;\;\;\; & \underline{{\cal H}} \\
\La \backslash {\bf C} \downarrow \;\;\;\;\;\;\;\;
& \;\;\;\;\;\;\;\;\;\;\;\;\;\;\;\; &
(2\pi i)^2 {\bf Z}\backslash {\bf C} \downarrow
\;\;\;\;\;\;\;\;\;\;\;\;\;\;\;\;\; \\
\D & \;\;\;\;\;\;\;\;\;\;\;\;\;\;\;\; &
{\bf P^1}\backslash \{ 0,1,\infty \}
\end{array}
\eeqa
where we {\em contrasted} with the Heisenberg bundle
(in principle we could also consider on the lhs the actual moduli
space after modding by the left integral shifts).

\newpage
\noindent
{\em \underline{Remark 1}: Metric corrections}

Euclidean D2-branes wrapped on ${\bf S_A^3}$ lead to
metric corrections
of the hypermultiplet moduli space [\ref{OV}], [\ref{SS}],
[\ref{GG}].
We relate this to a twisted  dilogarithm.
The corrections were
obtained approaching the conifold point $z\ra 0 $
in a suitable limit to keep the result independent of
the details of the embedding of the local conifold $T^*{\bf S^3}$
in a specific global Calabi-Yau model. The size $|w|=e^{-\rho}$
of the rescaled variable $w$ ($\la$ the string coupling)
\beqa
\frac{z}{\la}=:w=e^{i(\th +i\rho)}
\eeqa
was kept fixed.
With $x=\int_A C \, , t = \int_B C$ the metric (in the string frame) is
\beqa
ds^2=V^{-1}\Bigl( dt - (A_x dx
+ A_w d\bar{w}+A_{\bar{w}} dw)\Bigr)^2
+V(dx^2+dw d\bar{w})
\eeqa
where the scalar potential $V$ and the vector potential $A$ with
$\nabla V = \nabla \x A$ are
\beqa
V&=&\frac{1}{4\pi} \log \Bigl( \frac{\La^2}{|w|^2}\Bigr)
+ \frac{1}{2\pi} \sum_{n\neq 0}e^{2\pi i n x}K_0(2\pi |n| |w|)\\
A_x&=&-\frac{1}{2\pi} \th \, , \,
A_w = A_{\bar{w}}=\frac{1}{2\pi i}\sum_{n\neq 0}\sign (n)
e^{2\pi i n x}K_1(2\pi |n| |w|)
\eeqa
To make the instanton contributions manifest
expand the Bessel function $K_0$ (large $|w|$)
\beqa
V \sim \frac{1}{4\pi} \log \Bigl( \frac{\La^2}{|w|^2}\Bigr)
+ \frac{1}{\pi} \sum_{n> 0}\frac{1}{\sqrt{n |w|}}
\cos (2\pi n x)e^{-2\pi n |w|}
\eeqa
In the appropriate variable
(so the $n$-instanton action is $S_n=n|w|$)
(cf. the multiplet reduction discussed earlier, restricted to $Q$ one has
$\Om = \Re \, \Om = |\Om| = vol$)
\beqa
u_{\phi}:=e^{2\pi i \phi}=e^{-2\pi |w|+2\pi i x}
\;\; , \;\; \phi:= \int_Q C + i \, \frac{1}{\la} \Om = x+i\, |w|=x+ie^{-\rho}
\eeqa
this leads to an instanton contribution to the $d\rho^2$ component of
the metric as given by the following terms
in the effective action\footnote{keeping only the leading contribution
(not from $A_z$ expansion)
to the $n$-instanton term}
(up to a $1/2\pi$ normalization factor)
\beqa
\label{twisted Li}
 2\, \Re \, L(u_{\phi}) \, d\rho d\rho
=2\, \Re \Bigl( \sum_{n > 0} \frac{u_{\phi}^n}{n^2}\; S_n^{3/2}\Bigr)
\, d\rho d\rho
=4\, \frac{1}{|w|^{1/2}}\,
\Re \Bigl( \sum_{n > 0} \frac{u_{\phi}^n}{n^2}\; n^{3/2}\Bigr)
\, dw d\bar{w} + \dots
\eeqa
This shows a 'twisted' version
$L(u):=(-\frac{1}{2\pi})^{3/2}\sum_{n > 0} \frac{u^n}{n^2}\;
(n\Re \log u)^{3/2}$
of the superpotential $W=Li(u)$
from membrane instantons wrapped on ${\bf S^3_A}$ (cf. (\ref{Li})),
with the appropriate\footnote{\label{GG footnote}the extracted
universal power $S_n^{3/2}$ explains itself according to [\ref{GG}]
as $3/2=p+a$ where $p=2$ is the number of fields in the kinetic term
and $a=4(\frac{1}{4}-\frac{3}{8})$ should emerge form the Jacobian for
the change of variables from zero modes to collective coordinates
and a weight $1/4$ and $-3/8$ is associated to each bosonic and
fermionic collective coordinate, respectively (the hypermultiplet
contains four real scalars)}
power of $S_n$ extracted. As
one has the same form for the $dx^2$ component
recall that (\ref{metric component}) gave $2\Re S dc d\bar{c}$
and (\ref{IIA Heisenberg}) showed $S$ in a position analogous to $Li(e^c)$
(we had here the $e^{2\pi i x}$ in
$u_{\phi}$; still note the quaternionic vs. hyperk\"ahler difference).

\newpage

\noindent
{\em \underline{Remark 2}: The conifold monodromy}

Consider the moduli space
singularity caused by vanishing of the ${\bf S^3}$ of the
deformed conifold at $z=\vol({\bf S^3_A})\ra 0$
and its associated monodromy.
This is usually discussed in the framework of the type IIB
vector multiplet moduli space.
We wish to compare some of these type II(B)
relations with $M$-theory relations.

First, concerning
type IIB note that for the deformed conifold
$x^2+y^2+z^2+v^2=\mu$ with
$\Omega\sim \frac{dxdydz}{v}=\frac{dxdydz}{\sqrt{\mu -x^2-y^2-z^2}}$
and the three-cycles $A, B$ seen as ${\bf S^2}$'s
(spanned by $\Re \, y, \Re \, z$) over $x$ one has (cf. [\ref{CIV}])
\beqa
\label{CIV relation}
\int_{{\bf S^2}}\Omega \sim dx \sqrt{x^2-\mu}
\eeqa
and finds then
(the last equality up to ${\cal O}(\frac{1}{\sqrt[3]{c}})$ correction)
\beqa
\label{IIA coni}
z=\int_A \Omega
&=&\int_A \Re \, \Om + i \int_A \Im \, \Om =\int_A vol
= \frac{1}{2\pi i}\int_{-\sqrt{\mu}}^{\sqrt{\mu}}dx\sqrt{x^2-\mu}=\mu /4
\nonumber\\
F_z=\Pi=\int_B \Omega
&=&\frac{1}{2\pi i}\int_{\sqrt{\mu}}^{\sqrt{c}}dx\sqrt{x^2-\mu}
\approx \frac{1}{2\pi i}\Bigl( \frac{1}{2}c-z(\log c + 1 -\log z)\Bigr)
\eeqa
where $B\ra B-A$, or $\Pi \ra \Pi - z $, corresponds to
$c\ra e^{2\pi i}c$.

Now compare the $M$-theory relations. $Q\simeq A$ but, as remarked
already above, $D\not\simeq B$: the $D$ is an ${\bf S^3}$ {\em at}
some $r$, whereas $B\cong {\bf R^3}$ is the cone over ${\bf S^2}$
(which may be truncated with a cut-off at $c$). Usually the relations
to follow are considered [\ref{AW}], [\ref{C}]
at the semicalssical end $P_i$ corresponding to
$\int_Q\Y \approx \infty$ or $u_i\approx 0$. By contrast we
are now, in classical terms, near a phase transition where
$\vol(Q) \ra 0$ (i.e. one has reached the
'critical circle' $|u_i|=1$ [\ref{C}]),
so we are looking near $P_{i+1}$ but
still in variables properly adapted to $P_i$.
Nevertheless note that
(with $z_i:=\log u_i$)
\beqa
\label{M coni}
\zeta_i :=\Re \, z_i&=&\int_{Q_i} -\Upsilon
=-2\pi^2 r_0^3\nonumber\\
\xi_i:= \Re \, x_i&=& \int_{D_i}  vol \approx
r^3 + const \cdot f_i \, r_0^3
=r^3 + const' \cdot \Re \log y_i \, \Re z_i
\eeqa
Do they reflect an analogue
of $\Pi \sim z \log z$ in (\ref{IIA coni}) for $\int_B \Omega$
(turned off all $C$ field integrals) ?
Now note that at $P_{i}$ one has volume defects
$(f_{i-1}, f_i, f_{i+1})\sim \rho (1, -2, 1)$ with $\rho \ra \infty$
(cf app. \ref{cone app}, [\ref{AW}], [\ref{C}]).
Therefore $\Re \, \log y_i = f_i$ and
$\Re \, \log \eta_i = (f_{i-1}-f_{i+1})/3$ behave identically
near $P_{i+1}$. From this relation
$\Re \, \log y_i \sim \Re \, \log \eta_i$
and the interpretation $\be u_i \sim \eta_i$
(cf. [\ref{C}]) one has with
$\be u_i= \frac{1}{1-e^{z_i}}\approx \frac{1}{-z_i}$
that $\log z_i\sim \log y_i$ holds;
so (\ref{M coni}) gives indeed $\xi_i \sim \zeta_i \log \zeta_i $
(near $P_{i+1}$ we are at
$z_{(i)}\approx 0\leria u_i \approx 1 \leria  \be u_i \approx
\infty$, so according to
$y_i=\eta_{i+1}/\eta_{i-1}\approx 1/0 \approx \infty \approx
\eta_i$).

\newpage
\noindent
{\em \underline{Remark 3}:
Instanton interpretation of the Schwinger computation}\\
%\noindent
The classical non-supersymmetric Schwinger computation
has $N=2$ and $N=1$ avatars (cf. Introduction and
app. \ref{Schwinger})\footnote{\label{Schwinger footn}Note
that in the framework
of $N=2$ supersymmetry the original
non-supersymmetric Schwinger computation of a vacuum amplitude
(for the contribution of a scalar particle)
is transformed to a computation of an $R^2$ correction from
contributions of BPS states given by D0-branes
and wrapped D2-branes on an isolated ${\bf S^2}$, say,
(bound states of the lightest
states at strong coupling, corresponding to the $M$-theory 'lift'
where such a bound state corresponds to a M2-brane with momentum
around the ${\bf S^1_{11}}$)
as the spin content compensates for the extra insertions of the
curvature.}.
We point here to relations between even the {\em way of derivation}
of the membrane instanton superpotential
(\ref{Li}) and an {\em instanton reinterpretation} [\ref{KimPage}]
of the occurrence of $Li$ in the classical computation.
The latter gives with
%\footnote{to keep the formulae simple
%we stick to the case $e\, E=|e|\, |E|$}
\beqa
\tau = i\mu\;\;\;\;\;\; , \;\;\;\;\;\;
\mu = \frac{m^2}{2e}\frac{1}{E}
\eeqa
(we assume $e\, E=|e|\, |E|$)
for pair creation in a constant background electric field $E$
an imaginary energy density (giving the probability density)
(for background cf.
[\ref{Schwinger51}]
%, [\ref{Weisskopf}], [\ref{KimPage}],
%[\ref{SriniPadma}], [\ref{DunneHall}], [\ref{BSF}],
%[\ref{AdesiZerbini}], [\ref{GusynShovko}], [\ref{SoldatiSorbo}],
%[\ref{McArthurGargett}],
%[\ref{BlauVisserWipf}]
)
\beqa
\label{pair prod}
w\sim Li(e^{2\pi i\tau })
\eeqa

The one-loop effective action for electrons in a background
electro-magnetic field is
in the case of a static magnetic field
(with
$D\!\llap/ :=\ga^{\nu}(\pa_{\nu}+ieA_{\nu})$, $B_n$ the Bernoulli numbers)
\beqa
S_{eff}&=&-i \log \det ( i D\!\llap/ -m)=-\frac{i}{2}\log \det
(D\!\llap/^2+m^2)
=\frac{i}{2}\int_0^{\infty}\frac{ds}{s}
tr e^{-s(D\!\,\llap/^2+m^2)}\nonumber\\
&=&-\frac{e^2B^2}{8\pi^2}\int_0^{\infty}\frac{ds}{s^2}
(\coth s -\frac{1}{s}-\frac{s}{3})e^{-\frac{m^2}{eB}s}\nonumber\\
&=&\frac{e^2B^2}{2\pi^2}TV
\sum_{n=1}^{\infty}\frac{B_{2n+2}}{(2n+2)(2n+1)2n}
\Bigl( \frac{2eB}{m^2}\Bigr)^{2n}
\eeqa
(in the case of a static magnetic field this represents the
negative energy of the electrons in the background; the subtractions
of $\frac{1}{s}$ and $\frac{s}{3}$ correspond to the subtraction of
the zero-field contribution and a logarithmic charge renormalization,
respectively).

In the case of a static uniform electric field one gets (from $B\lra
iE$) a corresponding perturbative series but now the effective action
is {\em complex} with its imaginary part (a sum over non-perturbative
tunneling amplitudes; this comes from the poles which have moved onto
the contour of integration, the real part is the remaining regulated principal
part) giving the pair-production rate
\beqa
\label{effective action}
S_{eff}&=&\frac{e^2E^2}{8\pi^2}TV\int_0^{\infty}\frac{ds}{s^2}
(\cot s - \frac{1}{s} + \frac{s}{3})e^{-2\mu s}\nonumber\\
&=&\frac{e^2E^2}{8\pi^3}TV\Biggl(-4\pi
\sum_{n=1}^{\infty}\frac{B_{2n+2}}{(2n+2)(2n+1)2n}
\frac{1}{\tau^{2n}} + i Li(e^{2\pi i \tau})\Biggr)
\eeqa

It is instructive to compare the analogies to our situation.
Instead of the proper time formalism describing the Dirac
fermion in the constant uniform electric field
one can consider an equivalent one-dimensional static non-relativistic
harmonic oscillator [\ref{KimPage}].
In the (bosonic) Fourier-mode
$\Phi(t, \vec{x})=e^{i({\bf k_{\bot}}{\bf x_{\bot}}-\om t )}
\phi_{\om, {\bf k_{\bot}}}(x_{\|})$ one has for the
tunnel process wave function
$\phi_{\om, {\bf k_{\bot}}}(x_{\|})\sim D(a_{{\bf k_{\bot}}}, \xi)$
with $D$ a parabolic cylinder function and
\beqa
\xi=\sqrt{2/eE}\; x \;\;\; , \;\;\; x:=\om +eEx_{\|}\nonumber\\
a_{{\bf k_{\bot}}}=\frac{\mu}{2eE}\;\;\; , \;\;\;
\mu:=m^2+{\bf k_{\bot}}^2
\eeqa
The one-instanton action is an integral between the classical turning
points $x_{\pm}=\pm \sqrt{\mu}-\om$
\beqa
\label{Sk instanton action}
S_{{\bf k_{\bot}}}=i\int_{x_-}^{x_+}dx_{\|}\sqrt{x^2-\mu}
=\pi a_{{\bf k_{\bot}}}
\eeqa
Now the tunneling and the no pair-production probability for a fermion are
\beqa
P_{{\bf k_{\bot}}}^{tunnel}=e^{-2S_{{\bf k_{\bot}}}}\;\;\; , \;\;\;
P_{{\bf k_{\bot}}}^{no \; pair}=1-e^{-2S_{{\bf k_{\bot}}}}
\eeqa
With $<0, out|0,in>=e^{i\int d^4x \L}$ one has as vacuum-to-vacuum
transition amplitude $P_0$
\beqa
e^{-\; total \; pair \; production}
=\prod_{all\; states}P_{{\bf k_{\bot}}}^{no \; pair}
=|<0, out|0,in>|^2
=e^{-2VT\,\Im \L}
\eeqa
Here $\L$ is the effective Lagrangian: the dispersive real part
includes the non-linear terms in the Euler-Heisenberg Lagrangian, the
absorptive imaginary part describes the pair production $\Im S_{eff}$
in (\ref{effective action}).
For fermions to be produced $P_0$ should have a complex phase
so that $|P_0|^2<1$.
One gets as pair production rate per unit volume and unit time in
${\bf R}^{1,d}$
(using $\int d\om = eEV_{\|}$, $V=V_{\bot}V_{\|}$)
\beqa
\label{Schwinger production rate}
w&=&2\Im \L=\frac{1}{VT}\sum_{all\; states}
\log (1-e^{-2S_{{\bf k_{\bot}}}})
=\frac{2V_{\bot}}{V}\int\frac{d\om}{2\pi}
\frac{d{\bf k_{\bot}}^{d-1}}{(2\pi)^{d-1}}
\sum_{n=1}^{\infty}\frac{1}{n}e^{-2\pi n \mu}\nonumber\\
&=&\frac{2}{(2\pi)^d}
\sum_{n=1}^{\infty}\frac{eE}{n}\Bigl(
\frac{eE}{n}\Bigr)^{\frac{d-1}{2}}
e^{-\pi n m^2/eE}
=\frac{2 (eE)^{(d+1)/2}}{(2\pi)^d}Li_{(d+1)/2}(e^{\tau})
\eeqa
Now, the point of this remark is that the modulus and dilogarithm
even {\em arise} in this classical case, when interpreted
this way as an instanton calculation, the same way as in the
membrane instanton case. For notice
that the instanton action (\ref{Sk instanton action})
resembles the way $\vol({\bf S^3_Q})$ occurs in (\ref{CIV relation})
and that even the way $Li_2$ arises is identical; for the
difference between the determined integral over the momenta in
(\ref{Schwinger production rate}) and the undetermined integral
in $Li(e^x)=\int_{-\infty}^x dy\log \be e^y$ is only apparent (let $A=\pi/eE$):
$w\sim \int_0^{\infty} rdr\log ( 1-e^{-A(m^2+r^2)})
\sim \int_0^{\infty} d\rho \log (1-e^{-A(m^2+\rho)})
=\int_{m^2}^{\infty} d\si \log ( 1- e^{-A\si})
=\frac{1}{A}\int_{-\infty}^{-Am^2} d\chi \log ( 1- e^{\chi})
=-\frac{1}{A}Li(e^{2\pi i \tau})$.
The analogy should extend to a full
$\tau\in {\bf C}$.

\newpage
\section{\large\label{SeiWit}{\bf The flat bundle of five-dimensional
$N=2$ $SU(2)$ field theory}}
\resetcounter
The period $A_D$
%in the flat bundle
of 5D $N=2$ $SU(2)$ field theory
is given by $Li(e^A)$ (cf. (\ref{5D AD evaluation as Li})),
the 4D period $a_D$ relates to the conifold
(at the monopole point $\tilde{u}=1$)
\footnote{$\Xi_6$ the period of the $6$-cycle in type IIA,
or the vanishing ${\bf S^3}$ of the conifold in type IIB,
cf. (\ref{locking}), (\ref{periods});
for the string/$M$-theoretic embedding
of the 4D/5D field theories
cf. app. \ref{4D SW}, \ref{SW from IIA}}:
%\beqa
%\label{aD conifold}
$\Xi_6\sim a_D \sim x_+ \sim \tilde{u}-1$.
%\eeqa

4D $N=2$ gauge theories arise
from type IIA string theory on a Calabi-Yau $X$.
As type IIA is $M$-theory
reduced on a circle, the 4D gauge
theory is coming from a 5D gauge theory, given
by $M$-theory on $X\x {\bf S^1_R}$ with
$R=g_s^{2/3}l_{p,11}$.
Then a reinterpretation of the  4D (vector multiplet) prepotential
becomes possible. The latter consists of a piecewise cubic polynomial part
(in the scalar vev's of the vecor multiplets)
plus instanton corrections
(suppressed in the 5D decompactification limit). The world-sheet
instanton corrections are reinterpreted as one-loop corrections
in the 5D space-time ${\bf R^4}\x {\bf S^1_R}$.

The prepotential $\F_R$ of $M$-theory
on\footnote{The vector multiplets of the 5D gauge theory
given by $M$-theory on $X$ itself
are given (up to the graviphoton)
by sizes of two-cycles, the prepotential is
%\beqa
%\label{cubic form}
$\F =\frac{1}{6}\sum C_{ijk}T_iT_jT_k$
%\eeqa
(in a K\"ahler cone)
where $C_{ijk}$ are the classical
intersection numbers of the divisors $i,j,k$ and
the $T^i$ the areas of the dual two-cycles.}
$X \x {\bf S^1_R}$
has the cubic form as 5D limit ($R\ra \infty$)
and $\F_{4D}\sim \frac{1}{2}T^2 \Bigl( \log (\frac{T}{\La})^2
+\sum_{n=1}^{\infty}\frac{c_n}{(T/\La)^{4n}}\Bigr)$ as 4D limit
($R\ra 0$) which shows the instanton corrections\footnote{$T$ a
complex scalar of a 4D vector
built by a real scalar of a 5D vector and an ${\bf S^1}$ Wilson loop}.
$R\sim \epsilon \ra 0$ induces the scaling limit (\ref{scaling limit})
by $\La_{UV}\approx \frac{1}{R_{11}}=\frac{1}{g_sl_s}$ and
(\ref{size relations}).
Now even before taking the scaling limit one can interprete [\ref{NL}]
the world-sheet instanton sum
(\ref{gauge coupling from CY}) as a perturbative field-theoretic
one-loop correction in
${\bf R^4}\x {\bf S^1_R}$.\footnote{The Gromov-Witten invariants
correspond to contributions of 5D BPS states
(weighted according to
spin) given by membranes wrapped on holomorphic curves of classes
$nb+mf\in H_2({\cal N}(F_0), {\bf Z})$ of central charge
$Z_{n,m}\sim \frac{ nT_b+mT_f }{g_sl_s}$ and mass
$M_{n,m}\sim |Z_{n,m}|$;
$c_{n,m}$ is the Euler characteristic of the moduli
space ${\cal M}_{n,m}$, i.e. the 'number' of curves of degree
$(n,m)$ taking into account spin factors.}

To substantiate the interpretation take the weak
coupling limit $T_b\ra \infty$ (without $T_f\ra 0$); so
instantons wrapping the base decouple and with
$c_{0,m}=-2\delta_{m,1}$ one finds
\beqa
\label{gauge coupling weak limit}
\tau_f=2\log (1-e^{-T_f})
\eeqa
When $T_f\ra 0$ one gets with $\tau_f\sim 2\log M_W$
(from $T_f\sim M_W$)
%indeed
the expected gauge theory correction\footnote{perturbative
one-loop contribution; to get the
field theory instantons one needs the double scaling}.
However, before taking this limit $T_f\ra 0$ one finds
from (\ref{gauge coupling weak limit})
\beqa
\label{tau f}
\tau_f=- T_f + \log \sinh^2 \frac{T_f}{2}
\eeqa
(up to
%an additive constant
$2\log 2$; cf. (\ref{sinh relations})).
By $\frac{1}{g_4^2}\sim \frac{2\pi R}{g_5^2}$, i.e.
$\tau_5 \sim \frac{\tau_4}{2\pi R}$
this gives back the 5D term
(\ref{S1 corrections}) via $T_f=2\pi RM_W$
(up to the {\em linear piece},
the decompactification limit, cf. footn. \ref{alternative},
\ref{Vafa}).
\newpage
\subsection{\label{Li in 5D}The flat bundle
of the five-dimensional field theory}
At a generic point of the vector multiplet moduli space the
gauge group is broken to a product of
$U(1)$'s. The gauge coupling $\tau=\frac{4\pi i }{g^2}$ of such a
$U(1)$ gets one-loop corrections from particles charged under this
$U(1)$ (like the $W$-bosons)
which have in ${\bf R^5}$ the form
\beqa
e^2\int \frac{d^5k}{(k^2+M^2)^2}\sim e^2M
\eeqa
(up to a divergent constant; $e,M$
the charge and mass of the particle).
On ${\bf R^4} \x {\bf S}^1_R$ one has further corrections from
world-line instantons of the particle around the ${\bf S^1}$
\beqa
\label{S1 corrections}
M+2\sum_{k=1}^{\infty}\frac{1}{2\pi R k}e^{-2\pi R \, k M}=
\frac{1}{2\pi R}\log \sinh^2 \pi M R
\eeqa
with $2\pi R k M$ the action of a world-line k-instanton (properly
speaking one has to write $|M|$ as the mass which was in 5D the real
scalar component of a vector multiplet
has become complex after compactifying on a further
${\bf S^1}$ by combining with the Wilson loop).

Let us introduce besides the ordinary quantities $U, A, A_D$ of the
five-dimensional theory the following expressions which are rescaled
to dimensionless combinations
\beqa
\U=R^2 \, U \; , \; \A=RA \; , \; \A_D=RA_D \; , \; \zeta =\Lambda R
\eeqa
The $SU(2)$ curve for the Coulomb branch of the
$N=2$ theory on ${\bf R^4} \x {\bf S}^1_R$ is
([\ref{Nekrasov}] - [\ref{Eguchi Kanno}])
\beqa
y^2=(x^2-\La^4)(x-\frac{1}{2R^2}(\U^2 -1))
\eeqa
This corresponds\footnote{A quartic form
$y^2=(x^2+sx+1)^2-4\zeta^4$ is also used which corresponds to the quartic
form $\tilde{y}^2=(\tilde{x}^2-u)^2-\La^4$ of the 4D curve
(footn. \ref{quartic curve forms}) via
$\sqrt{2}R\tilde{x}=x+s/2\, , \, 2R^2\tilde{y}=y\, , \, -U=s/2$ (by
(\ref{uU relation}))}
with the four-dimensional Seiberg/Witten curve
(\ref{4D SW curve}) under
\beqa
\label{uU relation}
u=\frac{1}{2R^2} (\U^2-1)=\frac{1}{2}(R^2U^2-\frac{1}{R^2})
\eeqa
From the 4D relation $\Delta_4=4R^4(\Lambda^4-u^2)$
one finds for the 5D discriminant
\beqa
\Delta = (2\zeta)^2-(1-\U^2)^2 \;\;\;\;\;\;\;\;\;\;\;\;\;\;\;\;\;\;
\Bigl( \buildrel \Lambda \ra 0 \over \lra \;\;\;   -(1-\U^2)^2 \Bigr)
\eeqa
The relevant one-form is $\frac{d\la_5}{dU}=c\frac{dx}{y}$
(with $c=\sqrt{2}/(8\pi)$).
The period integrals of the curve
\beqa
\label{5D period integrals}
\frac{d A}{d U}= c\oint _{\al}\frac{dx}{y} \;\; , \;\;
\frac{d A_D}{d U}= c\oint _{\be}\frac{dx}{y}
\eeqa
 relate to the field-theoretical
quantities\footnote{one has the compatibility relations
$\frac{d\la_5}{dU}=\frac{d\la_4}{du}$ and
$\frac{d A}{d U}=\frac{d a}{d u} \; , \; \frac{d A_D}{d U}=\frac{d a_D}{d u}$}.
The effective coupling is given by
\beqa
\label{5Dcoupling}
\tau_5=\frac{dA_D}{dA}=\frac{dA_D/dU}{dA/dU}
\eeqa
In the weak coupling limit ($\U \approx \infty$, corresponding to the
4D weak coupling limit $u \approx \infty$)
one gets (cf. (\ref{Minf AD}))
(in $\tilde{\A_D}=2\A_D, \tilde{\A}=-4\A$
and for the monodromy around $\U \approx \infty$)
\beqa
\small{\left( \begin{array}{c}\tilde{\A_D}\\ \tilde{\A} \\ 1
\end{array} \right)}
%=\small{\left( \begin{array}{c}2\A_D\\ -4\A \\ 1 \end{array} \right)}
=\small{\left( \begin{array}{ccc}
                  1 &  \log \zeta &   0              \\
                  0 &      1      &   0              \\
                  0 &      0      &   1
                  \end{array} \right)   }
\small{\left( \begin{array}{c}Li(x)\\ \log x \\1\end{array}\right)}
\; , \;\;
\tilde{M}_{\infty}=  \small{\left( \begin{array}{ccc}
                  1 &   -2\cdot 2\pi i    & 2\pi i\log \zeta    \\
                  0 &      1      &    2\cdot 2\pi i      \\
                  0 &      0      &   1
                  \end{array} \right)}
\eeqa
Compare with the result that one has by (\ref{monodromy matrices})
with $M(l_{\infty})=M(l_0)^{-1}M(l_1)^{-1}$
\beqa
\label{Minf Li}
c_3=\small{\left( \begin{array}{c}Li(y)\\ \log y \\ 1 \end{array} \right)}
\;\;\;\; , \;\;\;\;
M(l_{\infty})=  \small{\left( \begin{array}{ccc}
                  1 &     2\pi i  &   0        \\
                  0 &      1      &  -2\pi i   \\
                  0 &      0      &   1
                  \end{array} \right)  }
\eeqa
So one has that the 5D analogue $\A_D$ of the conifold period $a_D$
\beqa
\label{5D AD evaluation as Li}
\tilde{\A_D}=Li(e^{\tilde{\A}})+ (\log \zeta) \tilde{\A}
\eeqa
is given by the dilogarithm\footnote{$Li(e^{\A})$ occurs in ${\cal A}_D$
as integral (\ref{Li relations}) of $Li_1(e^y)\sim \log (1-e^y)$,
i.e. $\approx \tau_5^{pert}$ by
(\ref{tau5pert}), (\ref{UA relations}),(\ref{sinh relations}).}
of (exp of) $\A$
up to a linear modification
(cf. [\ref{C}], sect. 4.4).

Note that the object
$\underline{A}:=
\tiny{\left( \begin{array}{c}A_D\\ A \\ m_R \end{array} \right)}$
with the mass scale $m_R:=1/R$ also has the monodromy (\ref{Minf AD})
around $U\approx \infty$. By contrast\footnote{Note the formal
relation [\ref{Kanno Ohta}] of
the 5D {\em massless} $N_f=1$ curve
$y^2=(x^2+sx+1)^2-c\zeta^3(x-1)$ ($c$ a constant which can be
gauged by rescalings) with
the 4D {\em massive}
$N_f=1$ curve $\tilde{y}^2=(\tilde{x}^2-u)^2-c\zeta^3(\tilde{x}+m)$
in the quartic form (footn. \ref{quartic curve forms})
for $m=\U-1=-\frac{s}{2}-1$ under $\tilde{x}=x+s/2\, , \,
u=(s/2)^2-1$.},
for the four-dimensional theory
with one massive hypermultiplet one has the
monodromy (\ref{4D massive monodromy}) around $a_0$.
\\ \\
\noindent
{\em computational details}

To get these results note that
the periods (\ref{5D period integrals})
fulfill the Picard-Fuchs
equation\footnote{\label{KO footn}cf. [\ref{Kanno Ohta}];
note also that using the relations
with the four-dimensional quantities this is equivalent to
$\pa_u [4(\Lambda^4-u^2)\pa^2_u-1]\Pi=0$ which shows in the bracket
the four-dimensional Picard-Fuchs operator}
\beqa
\label{PF}
\Biggl \lbrack \pa^2_{\U}
+\frac{1}{\U}\Bigl( \frac{4\U^2}{\Delta}(1-\U^2) -1 \Bigr)\pa_{\U}
-\frac{\U^2}{\Delta} \Biggr \rbrack \pa_{\U} \Pi=0
\eeqa
(where $\Pi=\oint_{\ga} \la_5$
%for a one-cycle $\ga$
).
(\ref{PF}) gives the Schwarzian differential equation
(cf. (\ref{Schwarzian})) for $\tau_5^{pert}$
\beqa
\{ \tau_5^{pert},\,\U\}=\frac{2\U^2}{(\U^2-1)^2}-\frac{3}{2}\, \frac{1}{\U^2}
\eeqa
(capturing$^{\ref{KO footn}}$ the perturbative part at large $\U$ by the
formal limit $\La \ra 0$)
with the solution
\beqa
\label{tau5pertfirst}
\tau_5^{pert}=\log (\U^2-1)
\eeqa
Reinserting the parameter $\zeta$ one gets the
$\tau_5^{pert}$ one is actually working with$^{\ref{KO footn}}$
(cf. below)
\beqa
\label{tau5pert}
\tau_5^{pert}=\log (\U^2-1)\; \lra \;
\log \Bigl( \frac{4}{\zeta^2}(\U^2-1)\Bigr)
\eeqa
corresponding naturally with $\tau_4^{\infty}\sim
\log 8 \frac{u}{\La^2}$ under (\ref{uU relation}).
One has an asymptotic relation
\beqa
\label{UA relations}
\A = \frac{1}{2}\log(\U+\sqrt{\U^2-1})\;\; , \;
\U = \cosh 2\A
\;\; , \;
\sqrt{\U^2-1}=\sinh 2\A
\; ,
\frac{d\A}{d\U} = \frac{1}{2} \, \frac{1}{\sqrt{\U^2-1}}
\eeqa
(like $u\approx a^2$ at infinity in 4D).
For the $\sinh$  occuring in (\ref{tau5pert})
recall that
\beqa
\label{sinh relations}
\log x+ \sum_{n\geq 1}\log (1+\frac{x^2}{n^2\pi^2})\; = \;
\log \sinh x \; = \; - (\log 2) + x +\log (1-e^{-2x})
\eeqa
This shows correctly the perturbative contribution of
the Kaluza-Klein tower
\beqa
\label{tau5pert development}
\tau_5^{pert}= \log \sinh^2 2\A = 2\log 2\A
+2\sum_{n\geq 1} \log ( 1+\frac{4{\cal A}^2}{(n\pi)^2})
\eeqa
(because, besides a divergent piece, the terms
$\log (A^2 + n^2 \frac{\pi^2}{4R^2})$ occur here).
Notice the following two limits of (\ref{tau5pert development}):
first one finds from $\tau_5^{pert}=2\log \sinh 2RA=2(2RA+\log
(1-e^{-4RA})-\log 2)$ that one has the logarithmic expression
$\tau_5^{pert} \ra 2\log 2A$ (up to a logarithmically divergent piece in $R$)
for the 4D limit of $R\ra 0$, as expected from the
beta-function in the Seiberg/Witten theory. Secondly, in the
5D decompactification limit $R\ra \infty$, one finds a
linear expression from\footnote{rescaling $\tau_5$ so that it
corresponds to the gauge coupling $\tau_5=1/g_5^2$
of the 5D theory; $\tau_5^{pert}$ and $\tau_f$ in
(\ref{tau5pert development}) and (\ref{tau f})
are in a sense 4D
as ${\bf S^1}$ is compact, (\ref{S1 corrections}) is 5D;
cf. the discussion after (\ref{tau f})}
$\tau_{5, decompact}^{pert}
=\lim_{R\ra \infty}\frac{\tau_5}{{\bf 2\pi i} R} \sim 2\cdot 2A$.

It will be useful to introduce a further equivalent variable
\beqa
z=e^{2\A}=\U+\sqrt{\U^2-1}\;\; , \; \U=\frac{z+1/z}{2}
\;\; , \;
\sqrt{\U^2-1}=\frac{z-1/z}{2}
 ,
\frac{d\U}{\sqrt{\U^2-1}}=d\log z
\eeqa
Actually the more precise relation for $\tau_5^{pert}$
is not (\ref{tau5pertfirst}), (\ref{tau5pert})
but\footnote{\label{alternative}cf. [\ref{Kanno Ohta}]; for a
corresponding analysis
based on (\ref{tau5pert}) cf. app. \ref{5Dother}
(cf. also remark after (\ref{tau f}))}
\beqa
\label{precise tau5pert}
\tau_5^{pert}=2\log (1-e^{-4\A})=\log 4\sinh^2 2\A   -4\A
=\log 4\frac{\U^2-1}{z^2}  \lra  \log \frac{4}{\zeta^2}\frac{\U^2-1}{z^2}
\eeqa
One gets
from (\ref{5Dcoupling}), (\ref{precise tau5pert})
for the dual period in the weak coupling limit of large $U$
\beqa
\label{ord eval}
\A_D &=& \frac{1}{2}\int \log \Bigl( \frac{4}{\zeta^2}\frac{\U^2-1}{z^2}\Bigr)
\frac{d\U}{\sqrt{\U^2-1}}
=\frac{1}{2}\Bigl(  \int \log \frac{4}{\zeta^2} d\log z
   +  \int \log (\frac{z-\frac{1}{z}}{2z})^2 \, d\log z \Bigr)\nonumber\\
&=&\frac{1}{2}\Bigl( Li(1/z^2)+\log \zeta\log(1/z^2)\Bigr)
 = \frac{1}{2}\Bigl( Li(e^{-4\A})-(\log \zeta) 4 \A\Bigr)
\eeqa
So one gets$^{\ref{alternative}}$
for the monodromy around $\U \approx \infty$
\beqa
\label{Minf AD}
\small{\left( \begin{array}{c}\A_D\\ \A \\ 1 \end{array} \right)}
=\small{\left( \begin{array}{c}
\frac{1}{2}Li(x)+\frac{1}{2}\log \zeta \log(x)\\
-\frac{1}{4}\log x \\ 1 \end{array} \right)}
\;\;\; , \;\;\;
M_{\infty}=  \small{\left( \begin{array}{ccc}
                  1 &   8\pi i    & 2\pi i\log \zeta    \\
                  0 &      1      &  -\pi i         \\
                  0 &      0      &   1
                  \end{array} \right) }
\eeqa

\newpage

\section{Discussion}

Although it may be already clear from the foregoing presentation
let us point to the direction in which we see the significance
of the mentioned relations. We compared a number of scenarios
where the dilogarithm, respectively
structures associated to it, occured as a quantum correction.
Given the markedly
special functional structure of the dilogarithm\footnote{Note that we
are not concerned with the {\em expansion} of expressions (like prepotentials
in $N=2$ theories) {\em in} polylog's but rather with the much more specific
case of the {\em pure} dilog.} (nobody would intend
to ask for an underlying pattern of different occurrences of the
sine function, for example) we think these similarities are worth
to be remarked in the first instance and worth to be asked
for being understood by some explanation.
The proper reason of the fact that
juxtapositions such as the ones above are possible should
be that the occurrence of the dilogarithm (or variants of it) as a
quantum correction in the examples mentioned has a common origin:
this may happen either by direct connections between the
respective scenarios or by some more general principle
(like for the logarithm in one-loop corrections).

For this reason we give a certain weight to the reformulation
of the transformation properties of the dilogarithm via the
upper triangular (Heisenberg) group. It is via this device
that we try to suggest a connection to other quantum corrections.
A well-known case concerns the Seiberg/Witten non-perturbative
solution of 4D $N=2$ $SU(2)$ gauge theory where logarithmic terms
near special points of the moduli space are captured
in the monodromy description by upper triangular (integral)
$2\x 2$ matrices. When enhanced with a further (mass) parameter
this description extends to similar $3\x 3$ matrices, as we recall
in (\ref{general monodromy with matter}); the precise form
(\ref{4D massive monodromy}) does not give the proper dilog
Heisenberg monodromy. By contrast, when one chooses for the
additional mass scale the KK-scale $m=1/R_5$ of 5D Seiberg/Witten
theory compactified on a circle of radius $R_5$, one gets indeed
the correct monodromy for the dilogarithm up to a specific
linear modification, as described in sect. 3 (we recall necessary
background on the relevant associated Schwarzian differential equation
and the corresponding Seiberg/Witten theory in app.'s
\ref{Special functions} and \ref{4D SW});
concerning the modification cf. also [\ref{C}], sect 4.4.
We propose that in this case the reason lying behind
this occurrence of the dilogarithm is the interpretation of the
dual period $a_D$ (resp. rather its five-dimensional incarnation)
as a conifold period after a stringy embedding of the gauge theory
(the necessary period relations are recalled in app. \ref{SW from IIA}).
So our reference scenario for the meaning of the dilogarithm is the
conifold (and the associated superpotential), cf. [\ref{vafa}] in type
IIA (resp. for the $M$-theory case [\ref{C}]).

Further, given the reformulation via the Heisenberg group,
another possible relation comes to mind:
besides the universal scenario given by considering (not an arbitrary
3-cycle in type IIA on a Calabi-Yau but) the local conifold limit
there is also the type IIA universal hypermultiplet.
For the conifold we think in the
technical description in terms of the $M$-theory lift and the
corresponding dilogarithm, cf. [\ref{C}]; this
has the advantage of changing the two sides of the transition
to two symmetrical ${\bf S^3}$'s. The universal hypermultiplet framework
has then firstly
a similar exchange symmetry in (\ref{involution}) resp. the corresponding
conjugation exchange in the ${12}$ and ${23}$ positions
in (\ref{IIA Heisenberg in e}).
Further these two universal structures compare via the
transformation behaviour
(\ref{IIA Heisenberg}), resp. (\ref{IIA Heisenberg in e}), and
(\ref{monodromy matrices}) and (\ref{flat section}); further
via the real invariant functions (\ref{invariant}) and (\ref{inv fct});
finally in light of the discussion after (\ref{membrane charges}).
It seems that the two real degrees of freedom embodied in
$C\sim c \epsilon_{ijk}$ (resp. for the conjugate polarization
$\epsilon_{\bar{i}\bar{j}\bar{k}}$) in the universal hypermultiplet
case correspond in some sense with the dual moduli related to the
${\bf S^3}$'s in the $M$-theory lift of the conifold; it should be
interesting to understand this further.

Yet another indication of such a connection comes from a different
perspective (cf. remark 1 of section 2).
After extracting an
appropriate power (conceptually justifiable
[\ref{GG}], cf. footn. \ref{GG footnote})
of the metric corrections in the hyperk\"ahler limit
[\ref{OV}] one
ends up with a correspondingly twisted version of the dilogarithm;
as discussed after (\ref{twisted Li}), such a metric term
fits nicely with the $2\, \Re S \, dc d\bar{c}$ in
(\ref{metric component}) and the proposed correspondences for the
quantities involved.
These observations point into the same direction as
above.\footnote{Another
parallelism of occurrences of the dilogarithm in differing situations
was noticed when the classical ($N=0$) Schwinger computation
was used [\ref{GV}] to derive corrections in ($N=1,2$) supersymmetric theories
(cf. Introduction, footn. \ref{Schwinger footn};
for associated background cf. app. \ref{Schwinger}). Besides
this known relation an instanton reformulation [\ref{KimPage}] of the
classical Schwinger computation is formally even more closely related
(remark 3, sect. 2)
to an (instanton) derivation of the dilogarithm.}

Clearly the most interesting question would be whether the occurrence
of the dilogarithm (or the Heisenberg group shifts)
in different scenarios is
universal as a consequence of certain physical principles,
like in logarithmic one-loop corrections of gauge couplings;
the theta-angle shifts responsible for that behaviour
(described by integral upper triangular $2\x 2 $ matrices)
may be now replaced with the more complicated shifts
(described by corresponding $3\x 3$ matrices)
related to two 'dual' objects like $\Om_{ijk}$ and
$\bar{\Om}_{\bar{i}\bar{j}\bar{k}}$ or ${\bf S^3_Q}$ and ${\bf S^3_D}$
(respectively the associated $c, \bar{c}$ or the corresponding
$M$-theory moduli).

If the way we presented things to make them amenable to a useful
comparison would have convinced the reader that such an underlying
principle might be reasonable to ask for and interesting to formulate
this paper would have fulfilled its task.

\newpage

\appendix

{\large {\bf Appendix}}

\section{\label{cone app}{\large {\bf The cone on ${\bf S^3}\x {\bf S^3}$}}}
\resetcounter

The cone on ${\bf S^3}\x {\bf S^3}$ has
the $G_2$ holonomy metric ($r \geq r_0$) [\ref{AW}], [\ref{C}]
\beqa
\label{metric}
ds^2=\frac{dr^2}{1-(\frac{r_0}{r})^3}
+\frac{r^2}{36}\Bigl( da^2+db^2+dc^2
-(\frac{r_0}{r})^3 (da^2-\frac{1}{2}db^2+dc^2) \Bigr)
\eeqa
Concerning metric perturbations which preserve $G_2$ holonomy one finds
(up to terms $y^2{\cal O}( (\frac{r_0}{y})^6 )$
in a new radial coordinate $y$) with $(f_1,f_2,f_3)=(1,-2,1)$
\beqa
\label{y metric}
ds^2=dy^2+\frac{y^2}{36}\Bigl( da^2+db^2+dc^2
-\frac{1}{2} (\frac{r_0}{y})^3 (f_1\, da^2+f_2\, db^2+f_3\, dc^2)\Bigr)
\eeqa
At small $r_0$ or large $y$ one finds the conical metric with the full
$\Sigma_3$ symmetry.
One has for the volume of $Q_i$ and
the $y$-dependent volume of $D_i$
embedded in $X_i$
(at large $y$)
\beqa
\label{vol Q}
\vol \, Q_i&= &2\pi^2 r_o^3\\
\label{vol D}
\vol \, D_i&=&\frac{2\pi^2}{27}y^3
\Bigl(1+\frac{3}{8}f_i (\frac{r_o}{y})^3
+{\cal O}( (\frac{r_0}{y})^6)\Bigr)
\approx \log c + \frac{1}{72} \, f_i \, \vol Q_i
\eeqa
Here, the first correction to the divergent piece
is the finite volume defect
$\frac{1}{72}\; f_i \; \vol({\bf S^3_{r_0}})$.

A holomorphic observable on ${\cal N}$ must combine
as SUSY partners
the $C$-field period
$\alpha_i=\int_{D_i}C$
with an order $1/r^3$ metric perturbation
(w.r.t. the conical metric), as in
\beqa
y_i=\exp\Bigl( kf_i+i(\alpha_{i+1}-\alpha_{i-1})\Bigr )
\eeqa
(with $\prod_i y_i=1$).
Actually one works with the quantity
(so $\eta_i=(y_{i-1}^2y_i)^{1/3},
y_i=\frac{\eta_{i+1}}{\eta_{i-1}}$)
\beqa
\label{etadef}
\eta_i=\exp\Bigl( \frac{k}{3}(f_{i-1}-f_{i+1})+i\alpha_i\Bigr )
\eeqa
$\eta_i$ is $0,1,\infty$ at $P_{i-1}, P_i, P_{i+1}$
where $P_i$ is the semiclassical end $\vol(Q_i)\approx \infty$.
One also has the membrane instanton amplitude as local coordinate $u_i$
at $P_i$ (vanishing there)
\beqa
\label{u eta variables}
u_i&=&\exp \Bigl( -T \mbox{vol}(Q_i)+i\int_{Q_i}C \Bigr )
\eeqa
We denote by $\Phi_i$ the physical modulus
to which $u_i$ is related via $u_j=e^{i\Phi_j}$
($Q_i$ is a supersymmetric cycle so $\Upsilon|_{Q_i}$ is the volume form),
i.e. $\Phi_j=\int_{Q_j}C+i\Upsilon=\phi_j+i\; \mbox{vol}(Q_j)$.

In a semiclassical regime with $D_i=0$ one has
$Q_i=\mp D_{i+1}$ so that
$\mp\int_{D_1}C =\int_{Q_3}C$, what is (with holomorphy) tantamount
to having $\eta_{i+1}\sim u_i$ to first order.
We assume that such a relation persists\footnote{Thereby a relation
$\eta_{i-1}=\frac{1}{1-\eta_i}$ holds also for the $u_i$:
$u_3=\be u_1=\frac{1}{1-u_1}$}
so that
\beqa
\label{ui etai relations}
\be u_i=\eta_i
\eeqa

\section{\label{UH mod sp}{\large {\bf The universal hypermultiplet
moduli space}}}
\resetcounter
One may parametrize the coset space $SU(2,1)/U(1)$
by two complex scalars $z_1$ and $z_2$ in the open
four-ball $|z_1|^2+|z_2|^2<1$ with K\"ahler potential
\beqa
\label{Kahl pot z1z2}
K = - \log (1 - |z_1|^2-|z_2|^2 )
\eeqa
and the Bergman metric with a left $SU(2,1)$ isometry group
\beqa
\label{bergman}
ds^2 = \frac{dz_1 d\bar{z}_1 + dz_2 d\bar{z}_2}
{1 - z_1\bar{z}_1-z_2\bar{z}_2}
+ \frac{(\bar{z}_1 dz_1 +\bar{z}_2 dz_2)(z_1 d\bar{z}_1 + z_2 d\bar{z}_2)}
{(1 - z_1\bar{z}_1-z_2\bar{z}_2)^2}
\eeqa
One may also switch to polar coordinates
($r<1, \th \in [0, \pi ) , \phi \in [0, 2\pi ), \psi  \in [0, 4\pi )$)
\beqa
\label{polar coord}
z_1 = r \cos \frac{\th}{2} e^{i \frac{\psi + \phi}{2}}
\;\;\;\; , \;\;\;\;
z_2 = r \sin \frac{\th}{2} e^{i \frac{\psi - \phi}{2}}
\eeqa
which gives with the $SU(2)$ one-forms
($d\si_i=-\frac{1}{2}\epsilon_{ijk}\si_j \we \si_k$)
\beqa
\si_1 &=&  \cos \psi \, d \th + \sin \psi \, \sin \th \, d \phi \nonumber\\
\si_2 &=& -\sin \psi \, d \th + \cos \psi \, \sin \th \, d \phi \nonumber\\
\si_3 &=& d\psi + \cos \th \, d\phi
\eeqa
and the complex vielbeins
\beqa
u = e^2 + i e^1 \;\;\;\; &,& \;\;\;\; v = e^r + i e^3\\
e^r = \frac{dr}{(1-r^2)} \;\;\;\; , \;\;\;\;
e^{1/2} &=& \frac{r}{2\sqrt{1-r^2}}\, \si_{1/2} \;\;\;\; , \;\;\;\;
e^3 = \frac{r}{2(1-r^2)}\, \si_3
\eeqa
for the K\"ahler metric
\beqa
ds^2 &=& \frac{dr^2}{(1-r^2)^2}
+ \frac{r^2}{4(1-r^2)}(\si_1^2 + \si_2^2)
+ \frac{r^2}{4(1-r^2)^2}(\si_3^2)\\
&=& u \, \bar{u} \; + \; v\, \bar{v} = (e^r)^2 + (e^1)^2 +(e^2)^2 +(e^3)^2
\eeqa
The relation to the variables $S$ and $c$ is given by
\beqa
z_1 = \frac{1-S}{1+S}\;\;\;\; , \;\;\;\; z_2 = \frac{2c}{1+S}
\;\;\;\;\;\; \;\;\;\;\;\;\;\;\;\;\;\;
S = \frac{1-z_1}{1+z_1}\;\;\;\; , \;\;\;\; c = \frac{z_2}{1+z_1}
\eeqa
and makes, via a K\"ahler transformation,
(\ref{Kahl pot z1z2}) equivalent to their
K\"ahler potential
\beqa
K = - \log (S+\bar{S} - 2c \, \bar{c})
\eeqa

The toroidal fibration (\ref{polar coord})
resp. (\ref{bundle comparison})
is analogous to the
fibration one gets in the case of signature
$SU(3)/U(2)\cong {\bf P^2_C}$ (where $e^{-K}=1+r^2$
instead of (\ref{Kahl pot z1z2})) [\ref{CKL}].

\section{\label{5Dother}{\large {\bf The alternative evaluation
in 5D gauge theory}}}
\resetcounter
There is an ambiguity whether to carry along the linear piece $\A$
in $\tau$ in (\ref{precise tau5pert})
(cf. remark after (\ref{tau f})).
Starting instead of (\ref{precise tau5pert}) from (\ref{tau5pert})
one gets
\beqa
\label{AD determination}
\A_D &=& \int \tau_5^{pert} d\A
=\frac{1}{2}\int
\log \Bigl( \frac{4}{\zeta^2}(\U^2-1)\Bigr)\frac{d\U}{\sqrt{\U^2-1}}\nonumber\\
%=\frac{1}{2}\Bigl( \int
%\frac{\log \Bigl( \frac{4}{\zeta^2}(\U^2-1)\Bigr)}
%{\sqrt{\U^2-1}} d\U \Bigr)\nonumber\\
&=&\frac{1}{2}\Bigl( Li(1/z^2)+\log \zeta\log(1/z^2)
+\log^2 z \Bigr)\nonumber\\
&=& \frac{1}{2}\Bigl( Li(e^{-4\A})-4(\log \zeta) \A
+4 \A^2\Bigr)
\eeqa
Note that in contrast to (\ref{ord eval}) here a quadratic term in $\A$
appears in $\A_D$.
Now for $-4{\cal A}=i\phi\in i{\bf R}$ being purely imaginary
(think of this special case like having a coupling constant
$\th + i\frac{4\pi}{g^2}$ of $\th =0$)
one obtains with (\ref{real part}) that this quadratic term vanishes
\beqa
\label{AD evaluation}
{\cal A}_D=\frac{1}{2}\Bigl( -( \frac{\pi i}{2}+\log \zeta)4{\cal A}
+\frac{\pi^2}{6}+i\Im Li(e^{-4{\cal A}})  \Bigr)
\eeqa
To understand this occurrence
of $\; \Im Li(e^{-4{\cal A}})=I(\phi)\;$ in ${\cal A}_D$
more directly compare (starting from
$\tau_5^{pert}= 2\log \frac{2}{\zeta}+2\log (-i \sin \frac{\phi}{2})$)
\beqa
\label{I AD comparison}
I(\phi)&=&-\int \log 2  \sin \frac{\phi}{2} d\phi
=-\Biggl( (\log 2) \phi
        +\int \log  \sin \frac{\phi}{2} d\phi\Biggr)\nonumber\\
\A_D&=&\int \tau_5^{pert} d\A
=-\frac{i}{2}(\log \frac{2}{\zeta}) \phi
-\frac{i}{2}\int \log (-i \sin \frac{\phi}{2})d\phi\nonumber\\
&=&-\frac{i}{2}\Biggl( (\log \frac{2}{\zeta} - \frac{i\pi}{2})\phi
+\int \log \sin \frac{\phi}{2}d\phi\Biggr)
%&=& -(\log \zeta) 2{\cal A} + \pi i \A -\frac{i}{2}\Im Li(e^{4{\cal A}})
\eeqa
where the last line of
(\ref{I AD comparison}) gives back
(\ref{AD evaluation}) up to an integration constant.
Now
\beqa
\label{c3 lin modif}
\left( \begin{array}{c}2\A_D\\ -4\A \\ 1 \end{array} \right)
=\left( \begin{array}{c}
Li(y)+ \log \zeta\log y +\frac{1}{4}\log^2 y \\
\log y \\ 1 \end{array} \right)\\
\label{Minfty lin modif}
M_{\infty}=\left( \begin{array}{ccc}
1&- 2\pi i & (2\log \zeta + 2\pi i)2\pi i \\
0 & 1 & 2\cdot 2\pi i \\
0 & 0 & 1
\end{array}\right)
\eeqa
(the quadratic term $\A^2$ of (\ref{AD determination}) caused by
the linear piece in $\tau$ can be seen also
in (\ref{c3 lin modif}), (\ref{Minfty lin modif})). The relations
${\tau_5^{pert}}^{\infty}=2\log \Bigl( \frac{2}{\zeta}\U\Bigr)$
and $\A^{\infty}=\frac{1}{2}\log 2\U$ give directly the asymptotic relation
$\A_D^{\infty}\approx
\frac{1}{2}(-4\A^{\infty} \log \zeta +4{{\cal A}^{\infty}}^2)$,
which is the limit case of (\ref{AD determination}).

\newpage

\section{\label{4D SW}{\large {\bf 4D $N=2$ $SU(2)$ field theory}}}
\resetcounter
Four-dimensional $N=2$ supersymmetric pure $SU(2)$ gauge theory
has as Seiberg/Witten curve [\ref{SW1}] the elliptic curve $E_u$
(varying over the $u$-plane) given by
\beqa
\label{4D SW curve}
y^2=(x^2-\La^4)(x-u)
\eeqa
The meromorphic one-form (with the constant $c=\frac{\sqrt{2}}{4\pi}$)
\beqa
\label{4D mero 1form}
\la_4=c(x-u)\frac{dx}{y}
\eeqa
leads to the periods ($\al , \be$ the one-cycles of $E_u$
over $[-\La^2, \La^2]$ and $[\La^2, u]$, respectively)
\beqa
\label{SW periods}
a=\oint_{\al} \la_4 \; , \; a_D=\oint_{\be} \la_4
\eeqa
where the dual parameter $a_D$ is the derivative of the prepotential $\F$
\beqa
\label{a_D}
a_D=\pa \F / \pa a
\eeqa
From $a$ and $a_D$ one gets the expression for the $N=1$ K\"ahler potential
\beqa
\label{SW Kaehler potential}
K= -\Im \, a \, \bar{a}_D
\eeqa
Further one has the relation [\ref{Matone}]
(for second equality [\ref{KKLMV}] cf. (\ref{IIA flux superpotential})
and sect. \ref{SW from IIA}, app.)
\beqa
\label{Matone relation}
\frac{1}{2\pi i}u=\frac{1}{4}(2\F-a\pa_a \F)=- \pa_S \F
\eeqa
When expressed in the rescaled dimensionless variable $\u:=\frac{u}{\La^2}$
the three special points are $\tilde{u}=\pm 1 , \infty$.
As the Legendre function $\la$ (cf. (\ref{ladef})) is gauged on the
three special points $0, 1, \infty$ let us switch the gauges
$( +1 , -1,  \infty)\lra (0, 1, \infty )$ which is effected by
$z\lra \underline{z}:=\frac{1-z}{2}$. One has then
\beqa
\label{u la relation}
\underline{\tilde{u}}(\tau):=\frac{1-\u (\tau)}{2}=\la(\tau)
\eeqa
Note that one has a certain (formal) triality symmetry
operating on the quantum moduli space
as one finds, with the identification (\ref{u la relation}),
that ${\bf P^1_{\underline{\tilde{u}}}}=\Gamma(2)\backslash {\bf H_2}$
has an $\Sigma_3$ action with quotient
$\Sigma_3 \back {\bf P^1_{\la(\tau)}}={\bf P^1_{j(\tau)}}$
where $j=\frac{27}{4}\prod_{i\in {\bf Z_3}}(\be^i \la - \be^{i+1}\la)$
(cf. (\ref{j from la})). Clearly in the actual physical interpretation
this $\Sigma_3$ symmetry is broken down to a ${\bf Z_2}$ symmetry
between $\u=+1$ and $\u=-1$.

One has further that $\frac{\pa}{\pa u} \la_4$ is equal to
$-\frac{c}{2}\frac{dx}{y}$, i.e. to the holomorphic one-form.
The $u$-derivatives of $a$ and $a_D$ give\footnote{Note that
the normalisations used ammount to $\tau(E_u)=\frac{da_D}{da}
=i\frac{F(\al\be^2 \underline{\tilde{u}})}{F(\be \underline{\tilde{u}})}
=\tau (\be \underline{\tilde{u}})$ with (\ref{hypergeom trafo prop}) and
(\ref{hypergeom la inversion}); cf. now
footn. \ref{la footnote}.}
thus the (holomorphic) periods $b_1$ and $b_2$ of the elliptic curve
(cf. [\ref{Kanno Ohta}]; here $F(z)=F_{\frac{1}{2}\frac{1}{2}1}(z)$
the hypergeometric function, cf. sect. \ref{Special functions})
\beqa
\frac{1}{\La}\frac{da}{d\u}
=\frac{1}{2}\sqrt{\be \underline{\tilde{u}}}
F(\be \underline{\tilde{u}})\;\;\;\;\;\; , \;\;\;\;\;\;
\frac{1}{\La}\frac{da_D}{d\u}=\frac{i}{2}F(\underline{\tilde{u}})
\eeqa
which fulfil the Picard-Fuchs equation
\beqa
\label{4D PF}
(1-\u^2)\pa^2_{\u} \Pi - \frac{1}{4}\Pi =0
\eeqa
{\em Inclusion of matter}

For the sake of comparison with results in the five-dimensional
theory let us consider also the case with matter.
First note that quite generally the monodromy has the following
block-upper-triangular form
\beqa
\label{general monodromy with matter}
\left( \begin{array}{c}a_D\\ a \\  1
\end{array} \right)
\lra \left( \begin{array}{cc}
                   N      &   v \\
                   0      &   1
                  \end{array} \right)
\left( \begin{array}{c}a_D\\ a \\  1
\end{array} \right)
\eeqa
where $N\in Sl(2, {\bf Z})$ and $v$ is an integral column vector.
We recall how this possibility (not realised in the pure
gauge theory) is actually realised in the theory with matter.

Recall first the central charge and mass of a BPS particle
in the {\em pure} gauge theory
\beqa
\label{central charge}
M=\sqrt{2}|Z| \;\;\; , \;\;\;
Z=n_e a + n_m a_D
\eeqa
By contrast, let us consider the case $N_f=1$
with one additional hypermultiplet as matter; this consists
of two chiral $N=1$ multiplets $Q$, $\tilde{Q}$ with $Q$ a quark
in the fundamental representation (for $SU(2)$ $\tilde{Q}$ will be again
in the fundamental representation). The global symmetry on the $u$-plane
(we are discussing the Coulomb branch) changes
from ${\bf Z_2}$ to ${\bf Z_3}$ (coming from the anomaly free subgroup
${\bf Z_{12}}$ of $U(1)_R$) and there are three singularities related
by this symmetry. Furthermore the instanton terms go in powers
of $\La_1^3$ (instead of $\La^4$) and terms with an odd number of
instantons vanish because of an anomalous "parity" ${\bf Z_2}$
in $O(2N_f)$ sending the quark $Q$ to the antiquark $\tilde{Q}$.
The curve is then $y^2=x^2(x-u)+t\La_1^6$ (with $t$ a constant which
can be absorbed in $\La_1$).

Now let us turn on an $N=2$ invariant mass term for the matter.
The curve is modified to
$y^2=x^2(x-u)+\La_1^3 (\frac{1}{4}m x +t\La_1^3)$
where $t=- \frac{1}{64}$
(for large $m$, with $\La_0^4=m\La_1^3$ held fixed,
this reduces to the curve\footnote{Note that in the
normalizations used in the pure gauge theory [\ref{SW1}]
one works with the
$\Ga(2)$ curve $y^2=(x^2-\La^4)(x-u)$ whereas in the framework of
general $N_f$ [\ref{SW2}] one works, even for $N_f=0$,
with the $\Ga_0(4)$ curve $y^2=x^2(x-u)+\frac{1}{4}\La^4x$.
\label{quartic curve forms}
Note further that besides these cubic forms of the representation of
the elliptic curve there are also corresponding quartic forms in use
($\Ga^0(4)$ curves; cf. for example [\ref{Klemm}]), namely
$y^2=(x^2-u)^2-\La_0^4$ and $y^2=(x^2-u)^2-\La_1^3(x+m)$ for $N_f=0$
and $1$, respectively.}
of the pure gauge theory).
The elementary particles (like electrons or quarks) would not
have $m_e=\sqrt{2}|a|$ but $m_e=\sqrt{2}|a + \frac{m}{\sqrt{2}}|$;
so when additional abelian conserved charges are present
they contribute to the central charge
and this modifies (\ref{central charge}) by the occurrence of the
$U(1)$ charge $S$ of the hypermultiplet to
\beqa
\label{central charge with matter}
Z=n_e a + n_m a_D + S \frac{m}{\sqrt{2}}
\eeqa
At $a_{\pm}=\pm \frac{m}{\sqrt{2}}$ one of the elementary quarks
becomes massless.
The behaviour at $a_+=a_0$
\beqa
a   &\approx & a_0\nonumber\\
a_D &\approx & c-\frac{i}{2\pi}(a-a_0)\log (a-a_0)
\eeqa
(with $c$ a constant)
implies for the monodromy around $a_0$
\beqa
\label{4D massive monodromy}
\left( \begin{array}{c}a_D\\ a \\  \frac{m}{\sqrt{2}}
\end{array} \right)
\lra \left( \begin{array}{ccc}
                  1 &      1      &   -1 \\
                  0 &      1      &   0  \\
                  0 &      0      &   1
                  \end{array} \right)
\left( \begin{array}{c}a_D\\ a \\  \frac{m}{\sqrt{2}}
\end{array} \right)
\eeqa
\\
\noindent
{\em Stringy realization of the $N=2 \lra N=1$ mass breaking}

The quantum corrected version $W=mu$
of the classical ($u=<tr \Phi^2>\approx a^2/2$)
mass deformation in the field theory
(providing mass to the scalar partner $\Phi$ of the vector multiplet
which leads to the breaking $N=2 \lra N=1$) is realised [\ref{TV}] as a flux
induced superpotential $W=\int_W \Om \we H_3$ in the type IIB string,
essentially because $u$ occurs according to (\ref{Matone relation}),
(\ref{periods}) among the Calabi-Yau periods. As near $\u=\pm 1$ a monopole
and a dyon become massless one gets by including the light states
\beqa
\label{locking}
W=mu+(a_D-a_0)\phi\tilde{\phi}
\eeqa
which leads by minimisation to monopole condensation $\phi\tilde{\phi}$
and locking on the monopole point $\u=\pm 1\llra a_D=a_0$. The stringy
realization proposed in [\ref{TV}] of this scenario started from the type
IIA superpotential (cf. sect. \ref{SW from IIA})
\beqa
\label{IIA flux superpotential}
W_{flux}\sim \int_X H_2\we t \we t \sim
(\int_{{\bf P^1_b}}H_2) \cdot \vol(K3)=n_{flux}\, \pa_s \F
\eeqa
where the type IIA periods $S$ and $\pa \F / \pa S$
correspond to $\vol({\bf P^1_b})$ and $\vol(K3)$.\footnote{Due to analytic continuation
(\ref{analytic continuation})
this has to be refined [\ref{CKLTh}]: actually
$W=mu\sim 2 i \Xi^2_{\infty} + \Xi^4_{\infty}=2it + \pa_s \F$.}

%Note that
%the prepotential $F(X^0, X^1, \dots , X^n)$ of the periods $X^i$
%is related to the prepotential $\F(t^1, \dots , t^n)$ of the K\"ahler
%coordinates $t^A=X^A/X^0$ via $F=(X^0)^2 \F$; thereby one finds the
%relation
%\beqa
%\label{X0 period relation}
%\frac{1}{X^0}\frac{\pa F}{\pa X^0}=2\F - \sum_a t^a \pa_a \F
%\eeqa

\newpage

\section{\label{SW from IIA}\large{{\bf 4D $N=2$ $SU(2)$ field theory
from IIA-theory}}}
\resetcounter
Let us reconsider the gauge theoretic descriptions from
the perspective of an
embedding into the type IIA string on a Calabi-Yau manifold.
There one has for the prepotential ($k$ counting the multiple covers)
\beqa
\F=\frac{1}{6}\sum C_{ijk}T_iT_jT_k + \, lower \, terms \,
  +\sum_{(d_i)\geq 0, k\geq 1}\frac{c_{(d_i)}}{k^3}\prod_i e^{-kd_iT_i}
\eeqa
with the $(d_i)$ indicating the degree of the primitive world-sheet
instantons. \\ \\
{\em A local Calabi-Yau model}

To become more
specific let us consider a local model for a $SU(2)$ singularity,
i.e. an $A_1$ fibration. This is given by a CY elliptically fibered
over one of the Hirzebruch surfaces $F_0, F_1, F_2$; we will consider
the case of $F_0={\bf P^1_b}\x {\bf P^1_f}$ in the limit of making the
size of the fibre large, i.e. we look only at the neighborhood of the
$F_0$ inside the CY which is then the local CY isomorphic to the
total space of the normal bundle of $F_0$ (inside the original CY)
given by the line bundle ${\cal O}(-2)\x {\cal O}(-2)$ over $F_0$.

We are interested in the $U(1)$ related to $T_f$ so let us note that
\beqa
\label{parameter identification}
\frac{1}{g_{YM}^2}\sim T_b \;\;\;\; , \;\;\;\;
m_W =T_f \cdot \frac{1}{g_sl_s}
\eeqa
as the $W$-boson will be represented by a D2-brane wrapped on
${\bf P^1_f}$. The relevant limits
are then the weak couling limit
$T_b\ra \infty$ and the singular limit $T_f\ra 0$ leading to enhanced
gauge symmetry. The gauge coupling for the $U(1)$ related to $T_f$ is
given by
\beqa
\label{gauge coupling from CY}
\tau_f=i\frac{\pa^2 \F}{\pa T_f^2}
=i\sum_{n,m\geq 0; k\geq 1} m^2 \frac{c_{n,m}}{k} q_b^{kn} q_f^{km}
\eeqa
($q_i=e^{-T_i}$).
The scaling limit [\ref{KKV}]
(which allows to extract the field-theoretic quantities)
\beqa
\label{scaling limit}
T_f\sim \epsilon \;\;\;\; , \;\;\;\; T_b\sim \log \frac{1}{\epsilon^4}
\eeqa
gives, assuming $c_{n,m}\sim \ga_n m^{4n-3}$
($m$ large, $n$ fixed), for the leading
($\epsilon$-independent) term the sum
of space-time instanton corrections of a gauge theory prepotential
%GIVE THIS IN GAUGE THEORY BEFORE
\beqa
\label{gauge coupling from scaling limit}
\tau_f\sim \sum_n \ga_n \Bigl( \frac{e^{-T_b}}{T_f^4}\Bigr)^n
\eeqa
For the scales note
$\La_{QCD}^4=\La_{UV}^4 e^{-\frac{1}{g_{UV}^2}}$ where
$\La_{UV}\approx \frac{1}{g_s}m_s=\frac{1}{g_sl_s}$:
the expansion parameter in (\ref{gauge coupling from scaling limit})
is $\frac{e^{-T_b}}{T_f^4}=\frac{e^{-T_b}}{(g_sl_sM_W)^4}$
by (\ref{parameter identification}), i.e.
$\frac{\La_{QCD}^4}{M_W^4}$ for the asserted
$\La_{UV}=\frac{1}{g_sl_s}$
\beqa
\label{size relations}
T_f=g_sl_sM_W\;\;\;\; , \;\;\;\;
e^{-T_b}=(g_sl_s)^4\La_{QCD}^4
\eeqa
\\
{\em A global Calabi-Yau model}

To make closer contact with the field theory it is useful to go to a
special global Calabi-Yau model, in our case of $SU(2)$ gauge theory
to the $ST$-Calabi-Yau
%\footnote{Type IIA string theory on $X_{ST}$
%is dual to a well-studied model of the heterotic string on $K3\x T^2$.}
$X=X_{ST}$ given by the hypersurface ${\bf P}_{11226}(12)$.
It is $K3$-fibered over
a ${\bf P^1}$ base and its two K\"ahler classes $S$ and $T$ measure
the complex volume of the base and the fibre, respectively.
One uses an integral symplectic basis for the periods in type IIB of its
mirror Calabi-Yau $W$, or equivalently a basis for $H_3(W, {\bf Z})$ at
the point of maximal unipotent monodromy (corresponding in $X$ to
the large volume limit). At this point $z^A=0$ one has a unique
analytic period, normalized as $X^0=1+\O(z)$, and $m=h^{2,1}(W)$
logarithmic periods $X^A$ which provide natural
special complex K\"ahler coordinates
$t^A=\frac{X^A}{X^0}=\frac{1}{2\pi i} \log z^A + \O(z)$.
The prepotential is homogeneous of degree two in the periods $X^I$
(with $q^A=e^{2\pi i t^A}$)
\beqa
F=&-  \frac{C_{ABC}}{3!}&\frac{X^AX^BX^C}{X^0}
+\frac{n_{AB}}{2}X^AX^B+c_AX^AX^0
-i\chi \frac{\zeta(3)}{2(2\pi)^3}(X^0)^2 + (X^0)^2 f(q)\nonumber\\
=&(X^0)^2 \F&=(X^0)^2 \Biggl( -\frac{C_{ABC}}{3!}t^At^Bt^C
+\frac{n_{AB}}{2}t^At^B +c_At^A -i\chi \frac{\zeta(3)}{2(2\pi)^3}+ f(q) \Biggr)
\eeqa
The $C_{ABC}=\int_{X}J_AJ_BJ_C$
are the classical intersection numbers in the type IIA interpretation
($J_A$ integral $(1,1)$ forms spanning the K\"ahler cone), and
$c_A=\frac{1}{24}\int_X c_2 J_A$; the world sheet instanton expansion
in $q$ is determined via mirror symmetry using type IIB on $W$.

This defines an integral basis for the periods
\beqa
\label{large complex structure periods}
\Xi_{\infty}=
\left( \begin{array}{c} X^0 \\ X^A \\
\frac{\pa F}{\pa X^A} \\ \frac{\pa F}{\pa X^0} \end{array} \right)
= X^0
\left( \begin{array}{c}  1 \\ t^A \\
\frac{\pa \F}{\pa t^A} \\ 2\F-t^A\frac{\pa F}{\pa t^A} \end{array} \right)
\eeqa
One has the leading order relations to the large complex structure
variables $z_t \propto e^{-T} \, , \, z_s \propto e^{-S}$ (we use also
the common rescaled variables $x=1728 z_t \, , \,
y=4z_s$). To make contact with the field theory one first
brings the relevant divisors in the two-dimensional moduli space
to generic intersections.
The order two tangency at $p=(x,y)=(1,0)$ between the conifold divisor
$\De_{con}=\{ \De_{con}^+ \De_{con}^- =(1-x)^2-yx^2=0 \}$
(with component divisors
$\De_{con}^{\pm}= \{ \Bigl( (1-x)\pm x\sqrt{y} \Bigr) \}$)
and the weak coupling divisor $\De_{weak}=\{ y=0 \}$ is resolved by
blowing up $p$ to the exceptional divisor $E\cong {\bf P^1}$ and
the divisors $\De_{con}^+, \De_{con}^- , \De_{weak}$ meet the
$u$-plane $E$ in the three special points $\u= +1, -1, \infty$.
Studying the {\em classical} field theory limit\footnote{where
classical gauge group enhancement to $SU(2)$ occurs at $a=0$; the
coordinate choice reflects the double scaling limit $\epsilon \ra 0$ in
$y=e^{-S}=\al'^2 \La^4 e^{-\hat{S}}=\epsilon^4$
and $1-x\sim \al' u = \epsilon^2 \u$ so that $l_s \La =\epsilon$; the
relation $m^2_{W_{\pm}}\sim e^{-S/2}\u$ (by $u_{weak}\sim
a^2$ and (\ref{central charge})) expresses
the $\frac{8\pi^2}{b_1 g^2(M_{str})} =-\log
\frac{m_{W_{\pm}}}{M_{str}}$ ($b_1=4$)}
at $\De_{weak}\cap E$
one chooses the coordinates $w_1=\frac{x\sqrt{y}}{1-x}=1/\u$ and
$w_2=1-x\sim \al' u $ (the latter up to higher corrections in $\al'$);
at the conifold branch $\De_{con}^+ \cap E$ one uses
$x_+=\frac{1-x}{x\sqrt{y}}-1=\u -1$ and $x_2=w_2$. One finds [\ref{CKLTh}]
as leading terms at $u=\infty$ and the monopole point,
respectively ($s=2\pi i S \propto \frac{2\pi i}{g^2}$, $y=e^{-S}$;
$\ka=\frac{i}{\pi} (\log 8 -1)$)
\beqa
\label{periods}
\begin{array}{c||c|c|c}
\Xi^A&\mbox{field theory}& \De_{weak}\cap E & \De_{con}^+ \cap E\\ \hline
\Xi^1&1             &1                               &1 \\
\Xi^2&\al' u        &w_2                             &x_2 \\
\Xi^3&\sqrt{\al'}a  &\frac{i}{\pi}\sqrt{w_2}         &
-\frac{\Xi^6}{2\pi i}\log x_+ + (1+\frac{i}{\pi}-2\ka)\Xi^6 \\
\Xi^4&s             &\frac{\Xi^1}{\pi i}\log (w_1w_2)&
\frac{\Xi^1}{\pi i}\log x_2 - \frac{1}{\pi i} \log (1-x_+)\\
\Xi^5&\al' u s      &\frac{\Xi^2}{\pi i}\log (w_1w_2)&
\frac{\Xi^2}{\pi i}\log x_2 - \frac{x_2}{\pi i} \log (1-x_+)\\
\Xi^6&\sqrt{\al'}a_D&\frac{\Xi^3}{\pi i}\log w_1 +\ka\Xi^3
& \frac{1}{\sqrt{2}\pi}\sqrt{x_2}x_+
\end{array}
\eeqa
The periods $\Xi$ in the field theory limit are related to the periods
$\Xi_{\infty}=(1,t,s; \pa_s \F, \pa_t \F, 2\F - s\pa_s \F - t\pa_t \F )$
in the large complex structure basis (\ref{large complex structure periods})
by\footnote{with
$A_{\pm}=(5\pi^4 \pm 12 \Ga^8(3/4))/(36\pi^4),
B=-2\pi^3 \sqrt{3}/\Ga^4(3/4)$ and the $r_i$ $\O(10)$ negative
real constants}
\beqa
\label{analytic continuation}
\small{\left(
\begin{array}{c} \Xi^1 \\ \Xi^2 \\ \Xi^3 \\ \Xi^4 \\   \Xi^5 \\  \Xi^6
\end{array} \right)}
=\small{\left(
\begin{array}{cccccc}
0 & 2iA_+B     & 0    & A_-B       & 0      & 0 \\
0 & 2iB        & 0    & B          & 0      & 0 \\
1 & 0          & 0    & -1/2       & 0      & 0 \\
0 & r_1+2iA_-B & 2A_-B& A_-B +ir_2 & -iA_+B & -iA_-B \\
0 & r_3+2iB    & 2B   & B + ir_4   & iB     & B \\
0 & 0          & 0    & 0          & 0      & 1
\end{array}\right)}
\small{\left(
\begin{array}{c} \Xi_{\infty}^1 \\ \Xi_{\infty}^2 \\ \Xi_{\infty}^3 \\
\Xi_{\infty}^4 \\   \Xi_{\infty}^5 \\  \Xi_{\infty}^6
\end{array} \right)}
\eeqa

\section{\large\label{Schwinger}
{\bf $N=2$ Schwinger computation}}
\resetcounter
For four-dimensional $N=2$ supersymmetric
compactification of the type IIA string on a Calabi-Yau $X$
the supermultiplet $W=F^+\cdots +\th^2 R^+$,
containing the self-dual parts of the gravi-photon field strength
and the curvature, respectively, leads in the four-dimensional effective
action to terms determined by just string $g$-loop contributions
[\ref{GV}]
\beqa
\int d^4x \, d^4\th \, F_g(t_i)(W^2)^g
=F_g(t_i)\int d^4x\, F_+^{2g-2}R_+^2
\eeqa
with the Kahler-moduli dependent functions $F_g$
(the genus $g$ topological partition function)
computable from world-sheet instantons
({\em holomorphic} maps $WS\ra CY$) as
\beqa
\label{Fg}
F_g=\sum_{holo}e^{-A(\Sigma_g)}
\eeqa
The Kahler-moduli are scalars in the vector multiplets
so the $F_g(t_i)$ are perturbatively and non-perturbatively exact
as the type II dilaton lies in a hypermultiplet.
Assuming the constant vev $<F^+>=\la$ one gets
as $R_+^2$ contribution
($F(\la):=\sum F_g \la^{2g-2}$)
[\ref{GV}]
\beqa
\Bigl \lbrack \sum_g F_g(t_i)\la^{2g-2} \Bigr \rbrack R_+^2
\eeqa
Besides the terms proportional to $\la^{2g-2}$ there will be
contributions to $R_+^2$ from terms like $e^{-1/\la}$, leading to the
full expression ${\cal F}_Z=\sum_g F_g \la^{2g-2} + {\cal O}(e^{-Z/\la})$
(with the BPS charge $Z$ of a particle
in the background field made explicit). Schwinger's computation gives
\beqa
\label{Schwinger computation}
{\cal F}_Z=\int_0^{\infty}\frac{ds}{s^3}(\frac{s/2}{\sinh s/2})^2e^{-Zs/\la}
\eeqa
Here one may wish to think of type IIB at the conifold where a
D3-brane wrapped on the vanishing ${\bf S^3}$ gives rise to a massless
hypermultiplet whose contribution to the $R^2$ term when it runs in the loop
is captured by this Schwinger one-loop
computation\footnote{Essentially
$\int_0^{\infty}\frac{ds}{s}Tr  \, e^{-s(\triangle +m^2)}
=\frac{1}{4}\int_0^{\infty}\frac{ds}{s} \frac{1}{\sinh^2 \frac{s}{2}}
e^{-s\frac{m^2}{eE}}$.
To preserve at least half of the supersymmetry
the background field has to be self-dual
(so $\vec{E}=\pm i \vec{B}$ in Minkowski space)
leading (for a boson) in the proper
time formalism to the one-loop determinant expression
%\beqa
$F(\vec{E},\vec{B},m)
=\frac{1}{2}Tr \log \det \Bigl( (i\pa -eA)-m^2\Bigr)
=\frac{e^2EB}{2\pi^2}\int_0^{\infty}\frac{ds}{s^3}
(\frac{s/2}{\sinh \frac{seE}{2}})(\frac{s/2}{\sin \frac{seB}{2}})$
%\eeqa
for the free energy
(with $E^2-B^2=
\vec{E}^2-\vec{B}^2$ and $EB=\vec{E}\cdot \vec{B}$;
at the lower integration bound one has to take a UV cut-off which in
string theory might be the string scale; this concerns the
first two terms in (\ref{Fpert coni}) which have a divergent
companion factor $\log \mu / \epsilon$). In the self-dual
case one gets
$f(\mu)=\frac{e^2E^2}{2\pi^2}\int_0^{\infty}\frac{ds}{s^3}e^{-is\mu}
(\frac{s/2}{\sinh \frac{s}{2}})^2=\frac{e^2E^2}{2\pi^2}{\cal F}(\mu)$.
(\ref{Fpert coni}) gives the higher polynomial
corrections to the Maxwell Lagrangian, i.e. the Euler-Heisenberg
Lagrangian.}
which integrates out the charged field to produce an effective action
whose real part is a polynomial in the even powers of the field strength.
With $i\mu=Z/\la$ one has for (\ref{Schwinger computation})
\beqa
\label{Fpert coni}
{\cal F}_{pert}=\sum_g\frac{B_g}{2g(2g-2)}\mu^{2-2g}
\eeqa
as perturbative part\footnote{The leading
contribution at large radius in (\ref{Fg})
comes from constant maps (where $\Sigma_g$
degenerated to a point) so the relevant moduli space
replacing the sum in (\ref{Fg}) is ${\cal M}_g \x X$ giving
$\frac{e(X)}{2} \int_{{\cal M}_g}c_{g-1}^3
=\frac{e(X)}{2} (-1)^{g-1}e({\cal M}_g)2\frac{\zeta(2g-2)}{(2\pi)^{2g-2}}
=\frac{e(X)}{2} \frac{B_g}{2g(2g-2)}\frac{B_{g-1}}{(2g-2)!}$ (the
Chern class referring to the Hodge bundle, given pointwise by the
holomorphic one-forms, over ${\cal M}_g$) as leading order
contribution of the ${\cal F}_g$ coefficient (cf. (\ref{Fpert coni}) for the
case of the conifold).}
from the large $\mu$ (small $\la$) expansion in inverse powers of
the dimensionless combination $\mu =\frac{m^2}{2eE}$ (in our BPS case
$m=e$)
and additional non-perturbative imaginary terms $e^{-2\pi n \mu}$.
One finds the absoprtive part (with $\tau = i \mu$, cf. (\ref{pair prod}))
\beqa
\Im {\cal F}(\mu)\sim Li(e^{2\pi i \tau})
\eeqa
describing the pair production of light $D3\bar{D3}$
brane-antibrane states.\footnote{The $R^2$ term
leads to $\chi$ and $\si$: the
action will contain terms
%\label{Euler signature}
$S=\dots + \frac{\chi +\frac{3}{2}\si}{2}F(\la, t_i)+
\frac{\chi -\frac{3}{2}\si}{2}\bar{F}(\bar{\la}, \bar{t}_i)$
so the perturbative
(real) part/imaginary part corrects the Euler character/signature
($\la=g_s(E+iB)$).}
[\ref{GV}]

\section{\label{Special functions}
\large {\bf Modular Forms and Triality symmetry}}
\resetcounter
{\em The theta functions $\th_2, \th_3,\th_4$ and the Eisenstein
series $E_k$}

The theta functions are defined by
their series developments in $q=e^{2\pi i \tau}$
\beqa
\th_2(\tau)=\sum_{n\in {\bf Z}}q^{\frac{1}{2}(n+\frac{1}{2})^2}
\;\;\; , \;\;\;
\th_3(\tau)=\sum_{n\in {\bf Z}}q^{\frac{1}{2}n^2}
\;\;\; , \;\;\;
\th_4(\tau)=\sum_{n\in {\bf Z}}(-1)^n q^{\frac{1}{2}n^2}
\eeqa
They have the modular transformation properties
\beqa
\label{theta trafo}
\th_2(\tau +1)&=&e^{\frac{2\pi i}{8}}\th_2(\tau)\;\; , \;\;
\th_2(-1/\tau)=(-i\tau)^{1/2}\th_4(\tau)\nonumber\\
\th_3(\tau +1)&=&\th_4(\tau)\;\;\;\;\;\;\;\; , \;\;
\th_3(-1/\tau)=(-i\tau)^{1/2}\th_3(\tau)\nonumber\\
\th_4(\tau +1)&=&\th_3(\tau)\;\;\;\;\;\;\;\; , \;\;
\th_4(-1/\tau)=(-i\tau)^{1/2}\th_2(\tau)
\eeqa
Because we will usually have to deal with their fourth powers we
introduce the notation
\beqa
A:=\th_2^4\;\;\;\; , \;\;\;\; -b:=B:=\th_3^4\;\;\;\; , \;\;\;\; C:=\th_4^4
\eeqa
These fulfill the relation $A+b+C=0$
(we introduced $b$ just to keep the ${\bf Z_3}$
symmetry).
Let us also introduce the Eisenstein series
and the absolute modular invariant $j$
(for which $j/12^3$ provides an isomorphism of
$\overline{Sl(2,{\bf Z})\backslash H}$
and ${\bf C^{\infty}}$,
mapping $\om, i, i\infty$ to $0, 1,\infty$)
\beqa
\label{E}
E_2(\tau)&=&1-24\sum_{n=0}^{\infty}\frac{nq^n}{1-q^n}
=\frac{24}{2\pi i}\pa_{\tau} \log \eta(\tau)\nonumber\\
E_4(\tau)&=&1+240\sum_{n=1}^{\infty}\frac{n^3q^n}{1-q^n}
=\frac{1}{2}(A^2+b^2+C^2)=-3 \si_2(e_i)\nonumber\\
E_6(\tau)&=&1-504\sum_{n=1}^{\infty}\frac{n^5q^n}{1-q^n}
=-\frac{1}{2}(A-b)(b-C)(C-A)=\frac{27}{2}\si_3(e_i)
=\frac{3}{2}\sum_ie_i^2\nonumber\\
E_4^3-E_6^2&=&12^3\eta^{24}=-\frac{27}{4}AbC\nonumber\\
%=-\frac{1}{4}a(e_i)
\label{j}
j(\tau)&=&\frac{E_4^3}{\eta^{24}}=12^3\frac{E_4^3}{E_4^3-E_6^2}
=\frac{1}{q}+744+196884q+\cdots
\eeqa
{\em The half-periods $e_i$}\\
\noindent
An important ${\bf Z_3}$ symmetric function set are the
half-periods\footnote{There are different conventions about
the $e_i$ in the literature; one finds also $e_2$ and $e_3$ interchanged.}
\beqa
\label{ei def}
e_1=\wp (\om_1/2)\;\;\;\; , \;\;\;\;
e_2=\wp ((\om_1+\om_2)/2)\;\;\;\; , \;\;\;\;
e_3=\wp (\om_2/2)
\eeqa
of the elliptic curve
$y^2=4x^3-g_2x-g_3=4(x-e_1)(x-e_2)(x-e_3)$
which are modular forms of weight two w.r.t. $\Gamma(2)$
and relate to the theta functions (with common $\sim$ factor)
%PROP $\frac{1}{3}\frac{\pi^2}{\om_1^2}$
\beqa
e_1 \sim C-b\;\;\;\; , \;\;\;\;
e_2 \sim A-C\;\;\;\; , \;\;\;\;
e_3 \sim b-A
\eeqa
%with the inversion relations TIMES $3\frac{\om_1^2}{\pi^2}$
%\beqa
%\label{De rel}
%A=\frac{e_2-e_3}{3}\;\;\;\; , \;\;\;\;
%b=\frac{e_3-e_1}{3}\;\;\;\; , \;\;\;\;
%C=\frac{e_1-e_2}{3}\;\;\;\; , \;\;\;\;
%\eeqa
The action of the triality group is
based on the
isomorphism\footnote{
$\tiny{\left(\begin{array}{cc}1&0\\0&1\end{array}\right)},
\tiny{\left(\begin{array}{cc}0&1\\-1&1\end{array}\right)},
\tiny{\left(\begin{array}{cc}1&-1\\1&0\end{array}\right)} $ and
$\tiny{\left(\begin{array}{cc}0&-1\\1&0\end{array}\right)},
\tiny{\left(\begin{array}{cc}1&-1\\0&1\end{array}\right)},
\tiny{\left(\begin{array}{cc}1&0\\-1&1\end{array}\right)}$
give representatives}
$\Sigma_3\cong Sl(2,{\bf Z})/\Gamma(2)$
(cf. (\ref{cross ratio isomorphism}), (\ref{Sl2 realisation})).
$A,b,C$ and the $e_i$ are cyclically permuted under
${\bf Z_3}$, $\be$ induces on the $e_id\tau$ the
permutation$^{\ref{permu action}}$
$P_{\be}$.
%=\tiny{\left(\begin{array}{ccc}1&2&3\\2&3&1\end{array}\right)}$.
One
has to adjust the actions on the upper half-plane
variable $\tau$ for the order two elements $\al \be^i$
(initially acting as in (\ref{Sl2 realisation})) involving
the inversion $\al: \tau \ra \frac{1}{\tau}$; to stay within
the {\em upper} half-plane\footnote{
$\tiny{\left(\begin{array}{cc}0&1\\1&0\end{array}\right)}
\not\in Sl(2,{\bf Z})$, and
$\Im \,
(\tiny{\left(\begin{array}{cc}a&b\\c&d\end{array}\right)}
\tau)
=\footnotesize{}\frac{1}{|c\tau +d|^2}\, \Im \, \tau$
gets an additional determinantal factor}
we define the adjusted operation
\beqa
\ga_{H} \tau= sign(\ga) \cdot \ga_{Sl(2)} \tau
\eeqa
The corresponding operations, such as $\tau \ra \frac{-1}{\tau}$, are the ones
which are actually induced from the $Sl(2,{\bf Z})$ action.
One has then (with (\ref{theta trafo})
and $d\, (\tiny{\left(\begin{array}{cc}a&b\\c&d\end{array}\right)}\tau)
=\frac{1}{(c\tau +d)^2}\; d\tau$)
\beqa
A \;d\tau \;\buildrel \tau\ra\tau +1 \over \lra -A \; d\tau \;\;\;\;, \;\;\;\;
b \;d\tau \;\buildrel \tau\ra\tau +1 \over \lra -C \; d\tau \;\;\;\;, \;\;\;\;
C \;d\tau \;\buildrel \tau\ra\tau +1 \over \lra -b \; d\tau
\\
A \;d\tau \;\buildrel \al_{H} \over \lra -C \; d\tau\;\;\;\;\;\; , \;\;\;\;\;\;
b \;d\tau \;\buildrel \al_{H} \over \lra -b \; d\tau\;\;\;\;\;\; , \;\;\;\;\;
C \;d\tau \;\buildrel \al_{H} \over \lra -A \; d\tau
\eeqa
and so $\al_{H}$ induces the permutation
$P_{\al_{H}}
=\tiny{\left(\begin{array}{ccc}1&2&3\\3&2&1\end{array}\right)}$, i.e.
%\beqa
%\label{eiinvers}
$e_1 d\tau \; |_{\al_{H} \tau }=e_3 d\tau ,
e_2 d\tau \; |_{\al_{H} \tau }=e_2 d\tau ,
e_3 d\tau \; |_{\al_{H} \tau }=e_1 d\tau $.
%\eeqa
So $\Sigma_3$ operates on
$e_id\tau$
via the permutation action\footnote{\label{permu action}
%\label{ei fund}
$
%P_{e}=\tiny{\left(\begin{array}{ccc}1&2&3\\1&2&3\end{array}\right)},
P_{\be}=\tiny{\left(\begin{array}{ccc}1&2&3\\2&3&1\end{array}\right)},
%P_{\be^2}=\tiny{\left(\begin{array}{ccc}1&2&3\\3&1&2\end{array}\right)};
P_{\al_{H}}=\tiny{\left(\begin{array}{ccc}1&2&3\\3&2&1\end{array}\right)},
P_{(\al\be)_{H}}=\tiny{\left(\begin{array}{ccc}1&2&3\\1&3&2\end{array}\right)},
P_{(\al\be^2)_{H}}
=\tiny{\left(\begin{array}{ccc}1&2&3\\2&1&3\end{array}\right)}$}
\beqa
e_i d\tau \; |_{\ga_{H} \tau }=e_{P_{\ga}(i)} d\tau
\eeqa
{\em The Legendre $\la$ function}

The Legendre function\footnote{\label{la footnote}Note
that here again conventions can differ, especially from
a different choice of the $e_i$. By
(\ref{cross ratio isomorphism})
the relations (\ref{latria}) can also be understood as
permutations in (\ref{ladef}) for the case $\infty, e_1, e_2, e_3$} is
defined as a cross ratio (cf. below)
of the half-periods
\beqa
\label{ladef}
%\la(\tau)=\frac{e_3-e_2}{e_1-e_2}=-\frac{A}{C}
\la(\tau)=\frac{e_2-e_3}{e_1-e_3}=-\frac{A}{b}
\eeqa
which is a modular function for $\Gamma(2)$ and transforms under
$\Sigma_3=Sl(2,{\bf Z})/\Gamma(2)$ as follows
\beqa
\label{la trafo explic}
\la(\tau)&=&\la(\tau)\;\;\;\; , \;\;\;\;
\la(\frac{1}{1-\tau})=\frac{1}{1-\la(\tau)}\;\;\;\; , \;\;\;\;
\la(\frac{\tau -1}{\tau})=\frac{\la(\tau) -1}{\la(\tau)}\nonumber\\
\la(-\frac{1}{\tau})&=&1-\la(\tau)\;\;\;\; , \;\;\;\;
\la(\tau -1)=\frac{\la(\tau)}{\la(\tau)-1}\;\;\;\; , \;\;\;\;
\la(\frac{\tau}{1-\tau})=\frac{1}{\la(\tau)}
\eeqa
i.e. $\la(\ga \tau)=
\ga \la(\tau)$ for $\ga \in {\bf Z_3}$ and $\la(\ga_{H} \tau)=
\ga \be \la(\tau)$ for $\ga \in \Sigma_3 \backslash {\bf Z_3}$.
For the ${\bf Z_3}$ transforms of $\la$
(generalising (\ref{ladef}))
and the absolute modular invariant
function $j(\tau)$
($\la$ provides an isomorphism of
$\overline{\Gamma(2)\backslash H}$ and ${\bf C^{\infty}}$,
mapping $i\infty, 0, 1$ to $0, 1,\infty$)
one has
\beqa
\label{latria}
%\la=-\frac{A}{C}=\frac{e_3-e_2}{e_1-e_2}\;\;\;\ , \;\;\;\;
%\be \la = -\frac{C}{b}=\frac{e_2-e_1}{e_3-e_1}\;\;\;\ , \;\;\;\
%\be^2 \la= -\frac{b}{A}=\frac{e_1-e_3}{e_2-e_3}
\la&=&-\frac{A}{b}=\frac{e_2-e_3}{e_1-e_3}\;\;\; , \;\;\;
\be \la = -\frac{b}{C}=\frac{e_3-e_1}{e_2-e_1}\;\;\; , \;\;\;
\be^2 \la= -\frac{C}{A}=\frac{e_1-e_2}{e_3-e_2}\\
\label{j from la}
\frac{3^3}{2^2}j&=&
(\la\; -\be \la)\cdot (\be \la -\be^2 \la)\cdot (\be^2 \la - \la )
=\frac{(\la+\om)^3(\la+\om^2)^3}{\la^2(1-\la)^2}
\eeqa
\newpage
The cross ratio
$z=cr_{z_1, z_2, z_3, z_4}=\frac{z_1-z_3}{z_1-z_4}/\frac{z_2-z_3}{z_2-z_4}$
of four points $z_1, z_2, z_3, z_4$ of ${\bf P^1(C)}$
gives, as one has the equalities
$cr_{z_1, z_2, z_3, z_4}=cr_{z_2, z_1, z_4, z_3}
=cr_{z_3, z_4, z_1, z_2}=cr_{z_4, z_3, z_2, z_1}$
but the index four subgroup $\Sigma_3$ operates effectively,
a realisation
of the isomorphism $Sl(2,{\bf Z})/\Gamma(2) \cong \Sigma_3 $
\beqa
\label{cross ratio isomorphism}
1 \lra V \rightarrow \Sigma_4 \lra
Sl(2,{\bf Z})/\Gamma(2) \cong \Sigma_3 \lra 1
\eeqa
considering a non-linear action of the triality symmetry
group $\Sigma_3$ on $z$
\beqa
\label{Sl2 realisation}
\begin{array}{ccccccc}
z & & & \be z = {\frac{1}{1-z}} & & & \be^2 z =\frac{z-1}{z} \\
\al z = \frac{1}{z} & & & \al \be z = 1-z & & & \al \be^2 z = \frac{z}{z-1}
\end{array}
\eeqa
{\em The hypergeometric function
$F_{\frac{1}{2}\frac{1}{2}1}$ and the Schwarzian}\\
\noindent
Gauss's hypergeometric function
$_2F_1(\al, \be ,\ga; z)=:F_{\al \be \ga}(z)$
solves the differential equation
\beqa
z(1-z)\frac{d^2F}{dz^2}+(\ga - (\al + \be +1)z)\frac{dF}{dz}-\al\be F=0
\eeqa
is given for $|z|<1$ by
$F_{\al \be \ga}(z)=\frac{\Ga (\ga)}{ \Ga (\al)\Ga (\be) }
\sum_{n=0}^{\infty} \frac{\Ga (\al +n) \Ga (\be +n)}{\Ga (\ga +n) }
\frac{z^n}{n!}$
and has the Euler integral
\beqa
F_{\al \be \ga}(z)=\frac{\Ga (\ga)}{ \Ga (\be)\Ga (\ga-\be) }
\int_0^1 t^{\be-1}(1-t)^{\ga-\be-1}(1-zt)^{-\al}dt
\eeqa
We are interested mainly in
$F(z):=F_{\frac{1}{2}\frac{1}{2}1}(z)$
which gives the inverse function to $\la$ by
\beqa
\label{hypergeom la inversion}
\tau(\la)=i\frac{F(1-\la)}{F(\la)}
\eeqa
The $\Sigma_3$ covariance properties (\ref{la trafo explic})
of $\la$ are equivalent to corresponding properties of $F$
\beqa
\label{hypergeom trafo prop}
F(\al \be^2 z)=\sqrt{\al \be z}F(z)\;\;\;\; , \;\;\;\;
F(\al z)=\sqrt{z} \Bigl( F(z)-i F(\al \be z) \Bigr)
\eeqa

For a function $z(x)$ one defines the Schwarzian derivative
\beqa
\label{Schwarzian}
\{ z , x \} =
\frac{z^{'''}}{z^{'}}-\frac{3}{2}\Bigl( \frac{z^{''}}{z^{'}}\Bigr)^2
\eeqa
which is $Sl(2, {\bf C})$ invariant
$\{ \frac{az+b}{cz+d} , x \} = \{ z , x \}$
(so a fractional linear
$z(x)$ solves $\{ z , x \} = 0$) and
has the composition rule
$\{ y , x \} = \{ y , z \} \Bigl( \frac{dz}{dx} \Bigr)^2 + \{ z , x \}$
(implying
the inversion $\{ x , z \} = -\Bigl( \frac{dx}{dz} \Bigr)^2 \{ z , x \}$).
For two independent solutions $y_1 \, , y_2$ of a second order
differential equation the quotient $z(x)=y_2(x)/y_1(x)$ satisfies
a Schwarzian differential equation
\beqa
\frac{d^2y}{dx^2} \, +
p(x) \, \frac{dy}{dx} \, + q(x) \, y =0 \;\;\;
\Longrightarrow  \;\;\;
\{ z , x \} = -\frac{p^2}{2}-p^{'} + 2q
\eeqa
{\em The dilogarithm}\\
\noindent
$Li=Li_2=\sum_{n\geq 1}x^n/n^2$ (cf. (\ref{Li repeat}) and [\ref{C}])
is member of a series of higher polylogarithms
($Li_1(x)=\sum_{n\geq 1}x^n/n=\log \be x\;\; , \;\;
Li_0(x)=\sum_{n\geq 1}x^n=x \cdot \be x$)
with a hierarchical relation
\beqa
\label{Li relations}
\frac{d}{dx}Li_{k+1}(e^x)=Li_k(e^x)
\eeqa
%Under the non-linear $\Sigma_3$ action
%(\ref{Sl2 realisation}) $Li$ has the following transformation behaviour
%\beqa
%\label{symrel}
%Li(\frac{1}{u})&=&-Li(u)-\frac{\pi^2}{6}-\frac{1}{2}\log^2(-u)\nonumber\\
%Li(1-u)&=&-Li(u)+\frac{\pi^2}{6}-\log u \log (1-u)
%\eeqa
We describe $Li$ on the critical circle $|u|=1$,
the boundary of the domain of convergence of the series
representation.
%\beqa
%\label{real imag decomposition}
%$Li(e^{i\phi})=\frac{\pi^2}{6}-\frac{1}{4}\phi(2\pi - \phi) +i I(\phi)$
%\eeqa
One has the elementary evaluation for the real part
\beqa
\label{real part}
\Re \; Li(e^{i\phi})=\sum_{n\geq 1}\frac{\cos n\phi}{n^2}
=\frac{\pi^2}{6}-\frac{1}{4}\phi(2\pi - \phi)
\eeqa
and the non-elementary odd function (of period $2\pi$) $I(\phi)$
for the imaginary part\footnote{in general one has to take the
absolute value of the expression the $\log$ of which occurs here}
\beqa
\label{I}
I(\phi):=\Im \; Li(e^{i\phi})=\sum_{n\geq 1}\frac{\sin n\phi}{n^2}=
-\int_0^{\phi}\log (2\sin \frac{\psi}{2})d\psi
\eeqa
Let us give the monodromies for the
dilogarithm.
Its differential equation\footnote{\label{Vafa}
An additional constant $\log \beta u_*$ in
$\frac{dT}{d \log u}=\log \beta u-\log \beta u_*$
would give $T(u)=Li(u) - \log \be u_* \log u$.} is
\beqa
\label{diffequ}
\frac{d\, Li}{d\, \log u}=\log \be u
\eeqa
The monodromy representation of the relevant fundamental group
$\pi_1 ({\bf P^1} \backslash  \{ 0,1,\infty \})$ (of loops based at
$1/2$, say)
describes the images of the generator loops
$l_i(t)$ ($i=0,1$, $t\in [0,1]$) which encircle (in the mathematically
positively oriented sense) $z=0$ and $z=1$, respectively
(then $l_{\infty} \circ l_1 \circ l_0=1$).
One has as multi-valuedness
\beqa
%\label{log multi valued}
\log z &\buildrel l_0 \over \lra &\log z + 2\pi i \nonumber\\
\label{multi valued}
\log \be z &\buildrel l_1 \over \lra & \log \be z -2\pi i \;\; , \;\;
Li(z) \buildrel l_1 \over \lra  Li(z) -2\pi i \log z
\eeqa
Now take as fundamental object a principal branch
(on $|z-1/2|< 1/2$) of
\beqa
\label{fundam solut}
L(z)=  \left( \begin{array}{ccc}
                  1 &  \log \be z &   Li(z)    \\
                  0 &      1      &   \log z   \\
                  0 &      0      &   1
                  \end{array} \right)
\eeqa
Analytic continuation along a loop $l$
in ${\bf P^1}\backslash \{ 0,1,\infty \}$ leads to $M(l)L(z)$ where
\beqa
M: \pi_1 ({\bf P^1} \backslash  \{ 0,1,\infty \}) \ra Gl(3,{\bf C})
\eeqa
defines the monodromy representation.
One finds the monodromy matrices (\ref{monodromy matrices}).
There are two equivalent ways to describe this:
in the {\em vector} resp. {\em Heisenberg picture}
one has the matrices $M(l_i)$ of (\ref{monodromy matrices}) resp.
the elements (\ref{heis mono}) expressing
the monodromies (\ref{multi valued})
for the vector $c_3$ of (\ref{monodromy matrices}) resp. the group
element (\ref{flat section}).

\section*{References}
\begin{enumerate}

\item
\label{AW}
M. Atiyah and E. Witten,
{\em "M-Theory Dynamics On A Manifold Of $G_2$ Holonomy"},
hep-th/0107177.

\item
\label{OV}
H. Ooguri and C. Vafa,
{\em "Summing up D-Instantons"},
Phys. Rev. Lett. {\bf 77} (1996) 3296,
hep-th/9608079.

\item
\label{SS}
N. Seiberg and S. Shenker,
{\em "Hypermultiplet Moduli Space and
String Compactification to Three Dimensions"},
Phys. Lett. {\bf B388} (1996) 521,
hep-th/9608086.

\item
\label{TV}
T.R. Taylor and C. Vafa,
{\em "RR Flux on Calabi-Yau and Partial Supersymmetry Breaking"},
Phys. Lett. {\bf B474} (2000) 130, hep-th/9912152.

\item
\label{CIV}
F. Cachazo, K. Intriligator and C. Vafa,
{\em "A Large N Duality via a Geometric Transition"},
Nucl. Phys. {\bf B 603} (2001) 3, hep-th/0103067.

\item
\label{AMV}
M. Atiyah, J. Maldacena and C. Vafa,
{\em "An M-theory Flop as a Large N Duality"}, hep-th/0011256.

%\item
%\label{AVI}
%M. Aganagic and C. Vafa,
%{\em Mirror Symmetry, D-Branes and Counting Holomorphic Discs},
%hep-th/0012041.

\item
\label{AV}
M. Aganagic and C. Vafa,
{\em "Mirror Symmetry and a $G_2$ Flop"}, hep-th/0105225.

\item
\label{C}
G. Curio,
{\em Superpotentials for $M$-theory on a $G_2$ holonomy manifold and
triality symmetry},
hep-th/0212211.

\item
\label{Joyce}
D. Joyce, {\em "On counting special Lagrangian homology 3-spheres"},
hep-th/9907013.

\item
\label{BGGG}
A. Brandhuber, J. Gomis, S.S. Gubser and  S. Gukov,
{\em "Gauge Theory at Large N and New $G_2$ Holonomy Metrics"},
Nucl.Phys. {\bf B611} (2001) 179,
hep-th/0106034.

\item
\label{Schwinger51}
J. Schwinger,
{\em "On Gauge Invariance and Vacuum Polarization"},
Phys. Rev. {\bf 82} (1951) 664.
\\
%\item
%\label{Weisskopf}
V. Weisskopf,
%{\em ""},
Kgl. Danske Videnskab. Selskabs. Met.-fys. Medd. {\bf 14} No. 6 (1936).
\\
%\item
%\label{SriniPadma}
K. Srinivasan and T. Padmanabhan,
{\em "Facets of Tunneling: Particle production in external fields"},
gr-qc/9807064;
{\em "Particle production and complex path analysis"},
Phys. Rev. {\bf D60} (1999) 024007,
gr-qc/9812028.
\\
%\item
%\label{DunneHall}
G.V. Dunne and Th.M. Hall,
{\em "Borel Summation of the Derivative Expansion and Effective
Actions"},
Phys. Rev. {\bf D60} (1999) 065002,
hep-th/9902064;
{\em "On the QED Effective Action in Time Dependent Electric
Backgrounds"},
Phys. Rev. {\bf D58} (1998) 105022,
hep-th/9807031.
\\
%\item
%\label{BSF}
A.B. Balantekin, J.E. Seger and S.H. Fricke,
{\em "Dynamical Effects in Pair Production by Electric Fields"},
Int. Jour. Mod. Phys. {\bf A6} No. 5 (1991) 695.
\\
%\item
%\label{AdesiZerbini}
V.B. Adesi and S. Zerbini,
{\em "Analytic continuation of the Hurwitz Zeta Function
with physical application"},
to be published in J. Math. Phys,
hep-th/0109136.
\\
%\item
%\label{GusynShovko}
V.P. Gusynin and I.A. Shovkovy,
{\em "Derivative Expansion of the Effective Action
for QED in 2+1 and 3+1 dimensions"},
J. Math. Phys. {\bf 40} (1999) 5406,
hep-th/9804143.
\\
%\item
%\label{SoldatiSorbo}
R. Soldati and L. Sorbo,
{\em "Effective action for Dirac spinors in the presence
of general uniform electromagnetic fields"},
Phys. Lett. {\em B426} (1998) 82,
hep-th/9802167.
\\
%\item
%\label{McArthurGargett}
I.N. McArthur and T.D. Gargett,
{\em "A ``Gaussian'' Approach to Computing Supersymmetric Effective Actions"},
Nucl. Phys. {\bf B497} (1997) 525,
hep-th/9705200.
\\
%\item
%\label{BlauVisserWipf}
S.K. Blau, M. Visser and A. Wipf,
{\em "Analytic Results for the Effective Action"},
Int. Jour. Mod. Phys {\bf A6} No. 30 (1991) 5409.

\item
\label{KimPage}
S.P. Kim and D.N. Page,
{\em "Schwinger Pair Production via Instantons in Strong Electric
Fields"},
to appear in Phys. Rev. D,
hep-th/0005078.

\item
\label{GG}
M.B. Green and M. Gutperle,
{\em "D-instanton partition functions"}.
Phys.Rev. {\bf D58} (1998) 046007,
hep-th/9804123.

\item
\label{BB}
K. Becker and M. Becker,
{\em "Instanton Action for Type II Hypermultiplets"},
Nucl. Phys. {\bf B551} (1999) 102,
hep-th/9901126.

\item
\label{GS}
M. Gutperle and M. Spalinski,
{\em "Supergravity Instantons and the Universal Hypermultiplet"},
JHEP {\bf 0006} (2000) 037,
hep-th/0005068 .

\item
\label{Ganor}
O.J. Ganor,
{\em "U-duality Twists and Possible Phase Transitions in 2+1D
Supergravity"}
Nucl. Phys. {\bf B549} (1999) 145,
hep-th/9812024.

\item
\label{Ketov}
S.V. Ketov,
%{\em "Gravitational dressing of D-instantons"},
%Phys. Lett. {\bf B504} (2001) 262,
%hep-th/0010255;
{\em "Universal hypermultiplet metrics"},
Nucl. Phys. {\bf B604} (2001) 256,
hep-th/0102099;
%{\em "Quantum geometry of the universal hypermultiplet"},
%talk RTN meeting, Corfu, September 2001,
%hep-th/0111080;
{\em "D-instantons and universal hypermultiplet"},
hep-th/0112012.

\item
\label{KKV}
S. Katz, A. Klemm and C. Vafa,
{\em "Geometric Engineering of Quantum Field Theories"},
Nucl. Phys. {\bf B497} (1997) 173, hep-th/9609239.

\item
\label{SW1}
N. Seiberg and E. Witten,
{\em "Monopole Condensation, And Confinement In N=2 Supersymmetric
Yang-Mills Theory"},
Nucl. Phys. {\bf B426} (1994) 19, Erratum-ibid. B430
(1994) 485, hep-th/9407087.

\item
\label{SW2}
N. Seiberg and E. Witten,
{\em "Monopoles, Duality and Chiral Symmetry Breaking in
N=2 Supersymmetric QCD"},
Nucl. Phys. {\bf B431} (1994) 484, hep-th/9408099.

\item
\label{Matone}
M. Matone,
{\em "Instantons and recursion relations in N=2 Susy gauge theory"},
Phys. Lett. {\bf B357} (1995) 342,
hep-th/9506102.

\item
\label{Klemm}
A. Klemm, {\em "On the Geometry behind N=2 Supersymmetric Effective
Actions in Four Dimensions"}, hep-th/9705131.

\item
\label{KKLMV}
S. Kachru, A. Klemm, W. Lerche, P. Mayr and C. Vafa,
{\em "Nonperturbative Results on the Point Particle Limit
of N=2 Heterotic String Compactifications"},
Nucl. Phys. {\bf B459} (1996) 537,
hep-th/9508155.

\item
\label{CKLTh}
G. Curio, A. Klemm, D. L\"ust and S. Theisen,
{\em "On the Vacuum Structure of Type II String Compactifications
on Calabi-Yau Spaces with H-Fluxes"},
Nucl. Phys. {\bf B609} (2001) 3, hep-th/0012213.

\item
\label{CKL}
G. Curio, B. Kors and D. L\"ust,
{\em "Fluxes and Branes in Type II Vacua and M-theory Geometry
with G(2) and Spin(7) Holonomy"},
hep-th/0111165.

\item
\label{Nekrasov}
N. Nekrasov,
{\em "Five Dimensional Gauge Theories and Relativistic Integrable Systems"},
Nucl. Phys. {\bf B531} (1998) 323, hep-th/9609219.

\item
\label{NL}
A. Lawrence and N. Nekrasov,
{\em "Instanton sums and five-dimensional gauge theories"},
Nucl. Phys. {\bf B513} (1998) 239, hep-th/9706025.

\item
\label{Kanno Ohta}
H. Kanno and Y. Ohta,
{\em "Picard-Fuchs Equation and Prepotential of Five Dimensional
SUSY Gauge Theory Compactified on a Circle"},
Nucl. Phys. {\bf B530} (1998) 73, hep-th/9801036.

\item
\label{Eguchi Kanno}
T. Eguchi and H. Kanno,
{\em "Five-Dimensional Gauge Theories and Local Mirror Symmetry"}
Nucl. Phys. {\bf B586} (2000) 331,  hep-th/0005008.

\item
\label{GV}
R. Gopakumar and C. Vafa,
{\em "M-Theory and Topological Strings--I, II"},
hep-th/9809187, 9812127.
{\em "Topological Gravity as Large N Topological Gauge Theory"},
hep-th/9802016, Adv. Theor. Math. Phys. {\bf 2} (1998) 413.
{\em "On the Gauge Theory/Geometry Correspondence"},
hep-th/9811131, Adv. Theor. Math. Phys. {\bf 3} (1999) 1415.

\item
\label{vafa}
C. Vafa,
{\em "Superstrings and Topological Strings at Large N"},
J.Math.Phys. 42 (2001) 2798,
hep-th/0008142.

\end{enumerate}
\end{document}